\documentclass[10pt]{article} 

\usepackage{hyperref}  
\hypersetup{
    colorlinks=true,      
    linkcolor=blue,       
    citecolor=blue,        
    filecolor=magenta,    
    urlcolor=blue         
}

\usepackage{fullpage}
\usepackage{enumitem}
\usepackage[page]{appendix}
\usepackage{amssymb}
\usepackage{mathrsfs} 
\usepackage{natbib}
\usepackage{algorithm}
\usepackage{algorithmic}
\usepackage{hyperref}
\usepackage{comment}
\usepackage{multirow}
\usepackage{amsthm}
\usepackage{bbm}
\usepackage{mathtools}
\usepackage{epstopdf,xcolor}
\usepackage{bbm}
\usepackage[colorinlistoftodos,prependcaption]{todonotes}
\usepackage{dsfont}
\usepackage{framed}
\usepackage{pifont}
\usepackage{subcaption}
\usepackage{tikz, subcaption}
\usetikzlibrary{shapes, arrows, positioning}
\usepackage{authblk}
\usepackage{listings}
\usepackage{tabularx}
\usepackage{booktabs,caption}
\usepackage[flushleft]{threeparttable}
\usepackage{longtable}

\newtheorem{theorem}{Theorem}

\newtheorem{proposition}{Proposition}

\newtheorem{definition}{Definition}

\newtheorem{remark}{Remark}

\newtheorem{assumption}{Assumption}

\newcommand{\E}{\mathbb{E}}

\newcommand{\independent}{\perp\mkern-9.5mu\perp}

\allowdisplaybreaks

\title{\bf \LARGE Treatment Policy Design in the Presence of Measurement Error\\~}

\author[1]{Chang Liu}
\author[1]{AmirEmad Ghassami\thanks{Corresponding author: AmirEmad Ghassami (email: \texttt{ghassami@bu.edu})}}

\affil[1]{Department of Mathematics and Statistics, Boston University, USA}

\date{}

\begin{document}

\maketitle

\begin{abstract}
In many applications, treatments are assigned based on unit features, leading to personalized treatment policies. This often involves optimizing objective functions with counterfactual quantities such as the conditional average treatment effect (CATE). In practice, some key features may be measured with error and ignoring such errors can introduce systematic bias. We study how to design personalized treatment policies when some features are latent and only noisy measurements are available. We ask how policies can be constructed when the CATE is not identifiable, and how uncertainty from measurement error can be incorporated into policy design. We propose two frameworks for treatment policy design based on partial identification, focusing on the measurement mechanism---the conditional distribution of observed measurements given unobserved features. The first framework builds on ideas from proximal causal inference. The second adopts a modular Bayesian approach based on a reparameterized likelihood, using Hamiltonian Monte Carlo and Bayesian regression models to separately perform posterior inference on the measurement mechanism and nuisance functions. Posterior quantiles are then used to construct credible intervals for the CATE that account for uncertainty in both components. Treatment policies are then derived by comparing the null value of the CATE to these bounds. The proposed methods provide a coherent way to account for measurement error and model uncertainty in personalized treatment policy design.

\end{abstract}

\section{Introduction}\label{sec:intro}
In many applications, decisions must be made about whether to assign a treatment or intervention to individuals, with the goal of improving their outcomes. Since individuals may differ in their characteristics and responses to treatment, personalized treatment policies aim to determine treatment assignments based on unit-level features. These policies are typically constructed by optimizing objective functions involving counterfactual quantities such as counterfactual risks, the average treatment effect (ATE), and the conditional average treatment effect (CATE) \citep{zhao2012estimating,zhang2012robust,dudk2014doubly,swaminathan2015counterfactual,luedtke2016statistical,athey2021policy,kallus2017recursive,zhou2017residual}. Most approaches are developed assuming unconfoundedness, under which the relevant value functions can be point identified from observed data. However, in many real-world settings, key unit features required for identification may be unavailable, only partially observed, or measured with error. In such cases, the relevant value functions may fail to be point identified. Ignoring missing or mismeasured features can therefore lead to biased estimation of treatment effects and inaccurate evaluations of the value function, resulting in suboptimal treatment policy design.

Existing research has developed various approaches to treatment policy design under unobserved confounding, including sensitivity analysis, instrumental variable methods, and proximal causal inference \citep{kallus2019interval, kallus2021minimax,cui2021semiparametric,qiu2021optimal,xu2021deep,shen2023optimal,qi2024proximal}. While these approaches provide important tools for addressing unobserved confounding, they often rely on additional identification assumptions or prespecified sensitivity models. In practice, such assumptions may be difficult to justify, particularly when the relationship between latent confounders and their observed proxies is not fully understood. This motivates the development of treatment policy design methods that accommodate uncertainty about the measurement mechanism and its implications for causal effect estimation.

Motivated by these challenges, in this paper, we investigate how to construct personalized treatment policies using error-prone measurements of latent confounders, while accounting for the uncertainty arising from the unknown measurement mechanism---the conditional distribution of observed measurements given unobserved features. In particular, we consider the setting where the underlying latent variable is discrete and its observed proxy is also discrete. This setup arises naturally in many real-world problems. For example, in medical settings, a patient's true disease severity category may not be directly observed, while only a physician-assigned stage or triage category is recorded. Such observed classifications are often subject to measurement error due to imperfect diagnostic tests, limited information at the time of assessment, or subjective clinical judgment \citep{balogh2015improving,young2018measurement}. We ask how treatment policies can be constructed when the CATE is not identifiable, and how uncertainty arising from measurement error can be incorporated into policy design. Under settings with discrete unobserved features, we formally establish the non-identifiability of the CATE. We then propose two novel frameworks for treatment policy design based on bound estimates, focusing on the measurement mechanism.

In Section \ref{sec:pi_proxy}, we introduce the first framework, which builds on ideas from proximal causal inference and derives conservative bounds on the CATE under mild invertibility conditions on the measurement mechanism. In Section \ref{sec:pc_mm}, we develop the second framework, which adopts a Bayesian approach based on a reparameterized likelihood to characterize posterior uncertainty in the measurement mechanism and nuisance functions, and constructs credible intervals for the CATE. We adopt a modular Bayesian approach \citep{bayarri2009modularization}, using Hamiltonian Monte Carlo to sample the measurement parameter and latent nuisance functions \citep{brooks2011handbook,hoffman2014no}, while separately estimating the observed-data nuisance functions through Bayesian regression models. We employ a Gaussian process prior to flexibly model the latent nuisance functions and use posterior quantiles of the CATE to construct credible intervals that account for uncertainty in both the measurement mechanism and the nuisance functions. In Section \ref{sec:modular_bayesian}, we establish theoretical properties of the modular Bayesian approach, providing justification for the proposed posterior inference procedure. The resulting intervals and bounds from the two frameworks are then used to construct personalized treatment policies in Section \ref{sec:policy_design}. In Section \ref{sec:sim},  we conduct simulation studies to evaluate the performance of the proposed treatment policies across different levels of measurement error, sample sizes, and prior specifications. We further assess the robustness of the proposed Bayesian framework under misspecified measurement mechanisms and compare the resulting policies with benchmark treatment regimes. Finally, Section \ref{sec:conc} summarizes our findings and provides concluding remarks.

\subsection{Related Work}
The challenges of treatment policy design under unobserved confounding have been studied from several perspectives, with different approaches addressing the lack of identification through additional assumptions, sensitivity analysis, or the use of auxiliary information. In this section, we review the literature most closely related to our work, including treatment policy design under unobserved confounding, Bayesian approaches to policy design under partial identification, and modular Bayesian inference for flexible posterior inference in complex statistical models.

A growing body of research has investigated treatment policy design when the unconfoundedness assumption is violated. \cite{kallus2019interval} proposed sensitivity-model-based methods that construct interval estimates for the CATE. Later, \cite{kallus2021minimax} developed a minimax regret framework for robust policy learning that optimizes treatment rules against worst-case violations of unconfoundedness within a sensitivity model. \cite{cui2021semiparametric} and \cite{qiu2021optimal} developed instrumental variable frameworks that leverage identification strategies under additional structural assumptions. \cite{xu2021deep} proposed a deep proxy causal learning approach for off-policy evaluation in a confounded bandit problem. \cite{shen2023optimal} and \cite{qi2024proximal} incorporated proximal causal inference into treatment rule learning by exploiting proxy variables for latent confounders to recover causal effects without relying on unconfoundedness assumptions. 

Bayesian approaches provide an alternative perspective on causal inference when unobserved confounding or measurement error prevents point identification. \cite{mccandless2007bayesian} developed a Bayesian sensitivity analysis for unmeasured confounding by incorporating prior information about the relationships involving latent confounders. \cite{gustafson2010simplified} further considered mismeasured and unobserved confounders, developing a simplified Bayesian sensitivity analysis to account for uncertainty arising from these variables. More generally, \cite{gustafson2015bayesian} provided a comprehensive overview of Bayesian inference under partial identification, emphasizing the role of prior information in characterizing posterior uncertainty about parameters that are not fully identified by the observed data.

As mentioned earlier, we utilized modular Bayesian inference in our proposed approach, which is a flexible approach to managing complex statistical models by separating posterior inference into distinct components, particularly when feedback between model components may be undesirable. \cite{bayarri2009modularization} discussed modularization in Bayesian analysis, emphasizing its role in reducing the influence of potentially misspecified model components. Related approaches based on composite likelihoods have been studied by \cite{pauli2011bayesian} and \cite{ribatet2012bayesian}, who investigated Bayesian inference using likelihood contributions constructed from lower-dimensional components. More recently, \cite{pompe2026asymptotics} studied the asymptotic properties of cut posterior distributions and developed robust modular inference procedures based on the posterior bootstrap. These developments provide a broader methodological foundation for separating posterior inference across different components of a statistical model.

\section{Problem Description}\label{sec:prob_dcrip}
Let $A\in\mathcal{A}=\{0,1\}$ denote a binary treatment, $Y\in\mathcal{Y}$ the observed outcome, and $X\in\mathcal{X}$ a vector of observed pre-treatment covariates. Throughout the paper, we assume that larger values of the outcomes $Y$ are preferable. Let $Y^{(a)}$ denote the potential outcome that would be observed if treatment were set to value $a$, and let $U\in\mathcal{U}$ represent unobserved confounders of the treatment--outcome relationship. We additionally have access to $W\in\mathcal{W}$, which is an error-prone measurement of $U$. We consider two types of measurement mechanisms defined as follows.

\begin{definition}
\label{def:non_differential} The measurement mechanism is called full-data non-differential if $W$ satisfies
\begin{align*}
W\independent\{Y^{(a)},A,X\}\mid U,\qquad a\in\{0,1\},
\end{align*}
and it is called observed-data non-differential if $W$ satisfies
\begin{align*}
W\independent\{Y,A,X\}\mid U.
\end{align*}

\end{definition}

We consider the setting in which $W$ and $U$ are both categorical with number of levels $k_w$ and $k_u$, such that $W$ and $U$ take values in $\{1,
\dots,k_w\}$ and $\{1,
\dots,k_u\}$, respectively. Throughout the paper, we assume $k_w\geq k_u$. In our setting, we observe independent and identically distributed samples $\mathcal{D}_n=\{A_i,X_i,W_i,Y_i\}_{i=1}^n$ drawn from the target distribution.

\subsection{Treatment Regimes and Treatment Policy Design}\label{sec:reg_policy_design}
A treatment regime $g$ is a deterministic decision rule mapping the covariate (and possibly measurement) space to the treatment space $\mathcal{A}$. Under a regime $g$, let $Y^{(g)}$ denote the corresponding potential outcome. We consider the following two types of optimal regimes depending on the information available for decision making.

\begin{definition}[$X$-optimal regime]\label{def:x-opt} The $X$-optimal regime $g_{\text{opt}-X}$ assigns treatment $a$ given $x\in\mathcal{X}$ as
\begin{align*}
    g_{\text{opt}-X}(x):={\arg\max}_{a\in\{0,1\}}\mu_a(x),
\end{align*}
where $\mu_a(x):=\mathbb{E}[Y^{(a)}\mid X=x]$.
\end{definition}

\begin{definition}[$(X,W)$-optimal regime]\label{def:xw-opt} The $(X,W)$-optimal regime $g_{\text{opt-}(X,W)}$ assigns treatment $a$ given $x\in\mathcal{X}$ and $w\in\{1,\dots,k_w\}$ as
\begin{align*}
    g_{\text{opt-}(X,W)}(x,w):={\arg\max}_{a\in\{0,1\}}\mu_{a,w}(x),
\end{align*}
where $\mu_{a,w}(x):=\mathbb{E}[Y^{(a)}\mid X=x,W=w]$.
\end{definition}

For comparison, we also introduce a third regime of interest, namely the naive regime, which ignores the latent structure.
\begin{definition}[$(X,W)$-naive regime]\label{def:xw-nai} The $(X,W)$-naive regime $g_{\text{naive-}(X,W)}$ assigns treatment $a$ given $x\in\mathcal{X}$ and $w\in\{1,\dots,k_w\}$ as
\begin{align*}
    g_{\text{naive-}(X,W)}(x,w):={\arg\max}_{a\in\{0,1\}}m_{a,w}(x),
\end{align*}
where $m_{a,w}(x):=\mathbb{E}[Y\mid A=a,X=x,W=w]$.
\end{definition}

The following proposition states that $(X,W)$-optimal regime is always better than, or at least as good as $X$-optimal regime and $(X,W)$-naive regime in terms of potential outcome means.
\begin{proposition}\label{prop:xw_better} The $(X,W)$-optimal regime
\begin{itemize}
    \item[(a)] outperforms the $X$-optimal regime in the sense that,
    \begin{align*}
        \mathbb{E}[Y^{(g_{\text{opt}-(X,W)})}]\geq\mathbb{E}[Y^{(g_{\text{opt}-X})}],
    \end{align*}
    \item[(b)] outperforms the $(X,W)$-naive regime in the sense that,
    \begin{align*}
        \mathbb{E}[Y^{(g_{\text{opt-}(X,W)})}]\geq\mathbb{E}[Y^{(g_{\text{naive-}(X,W)})}].
    \end{align*}
\end{itemize}
    
\end{proposition}

Proposition \ref{prop:xw_better}(a) follows from enlarging the information set used for treatment assignment, i.e., the regime $g_{\text{opt-}(X,W)}$ uses more observed information compared to $g_{\text{opt-}X}$, as it is optimized not only with respect to $X$ but also with respect to $W$. Proposition \ref{prop:xw_better}(b) does not rely on enlarging the information set, as both regimes are functions of $(X,W)$. Instead, it follows from the fact that $g_{\text{naive-}(X,W)}$ is constructed by optimizing a misspecified value function that treats $W$ as a sufficient adjustment variable and ignores residual confounding through the latent variable $U$. Consequently, $g_{\text{naive-}(X,W)}$ need not maximize the true potential outcome value function $\E[Y^{(g)}]$, whereas $g_{\text{opt-}(X,W)}$ is defined to do so.

Although $g_{\text{opt-}(X,W)}$ is population-level optimal relative to the other two regimes, its identification presumes knowledge of the corresponding conditional mean function $\mu_{a,w}(x)$. In the presence of the latent confounder $U$, however, neither $\mu_{a,w}(x)$ nor $\mu_a(x)$ is in general point identified, as established in Section~\ref{sec:identifiability}. Consequently, neither the $(X,W)$-based optimal regime $g_{\text{opt-}(X,W)}$ nor the $X$-based optimal regime $g_{\text{opt-}X}$ can be directly recovered from the observed data. This lack of point identification motivates finding bounds on the conditional mean functions to inform treatment assignment.

\begin{remark}
It is important to distinguish between bounding the conditional mean functions and bounding the conditional average treatment effect (CATE). As mentioned by \citet{Laurendeau02012026}, identification of $g_{\text{opt-}(X,W)}$, $g_{\text{opt-}X}$, or $g_{\text{naive-}(X,W)}$ does not require point identification of the corresponding conditional mean functions. Instead, it is sufficient to determine the sign of the relevant CATE. Specifically, define the $(X,W)$-CATE,
\begin{align*}
    \tau(x,w)&:=\mathbb{E}[Y^{(1)}-Y^{(0)}\mid X=x,W=w]=\mu_{1,w}(x)-\mu_{0,w}(x),
\end{align*}
the $X$-CATE,
\begin{align*}
    \tau(x)&:=\mathbb{E}[Y^{(1)}-Y^{(0)}\mid X=x]=\mu_1(x)-\mu_0(x),
\end{align*}
and the naive $(X,W)$-CATE,
\begin{align*}
    \tau_{\mathrm{naive}}(x,w)&:=\mathbb{E}[Y\mid A=1,X=x,W=w]-\mathbb{E}[Y\mid A=0,X=x,W=w]=m_{1,w}(x)-m_{0,w}(x).
\end{align*}
If the sign of the corresponding CATE is identified, then the associated optimal regime is identified, since treatment is assigned whenever the CATE is positive.

The observed-data conditional mean functions $m_{a,w}(x)$ are directly identifiable from the observed data. In contrast, the causal conditional mean functions $\mu_{a,w}(x)$ and $\mu_a(x)$ are generally not point identified under measurement error, motivating the need for bounds. Bounds on the conditional mean functions immediately induce bounds on the corresponding CATE by taking differences. Therefore, in Section~\ref{sec:pi_proxy}, we focus on partial identification of the conditional mean functions and obtain CATE bounds through this construction. In Section~\ref{sec:pc_mm}, we adopt a Bayesian approach and directly derive posterior credible intervals for the CATE.

Finally, even if the CATE itself is bounded, its sign may remain unidentified whenever the resulting bounds contain zero. In such cases, the optimal treatment decision cannot be uniquely determined from the available information, and treatment recommendations must account for this remaining uncertainty.

\end{remark}

\subsection{Identifiability under Latent Confounding}\label{sec:identifiability}

In this section, we study the identification properties of the conditional mean functions $\mu_{a,w}(x)$ for $a\in\{0,1\}$, for all $x\in\mathcal{X}$ and $w\in\{1,\dots,k_w\}$. We begin by introducing the following standard causal identifiability assumptions.

\begin{assumption}[Latent unconfoundedness]\label{ass:latent_uncon}
\begin{align*}
    Y^{(a)}\independent A\mid\{U,X\},\quad\text{for }a\in\{0,1\}.
\end{align*}
\end{assumption}

\begin{assumption}[Consistency] \label{ass:consistency}$Y=Y^{(a)}$ when $A=a$ almost surely for $a\in\{0,1\}$.
\end{assumption}

\begin{assumption}[Positivity] \label{ass:positivity}$p(a\mid X,U)>0$ almost surely for $a\in\{0,1\}$.
\end{assumption}

\begin{remark}
    Assumption \ref{ass:latent_uncon} is a much weaker version of standard unconfoundedness assumption, as it only requires conditional independence after adjusting for the latent confounder $U$ in addition to $X$.
\end{remark}

In the presence of latent confounding through $U$, $\mu_{a,w}(x)$ is not identifiable from the observed data, even after imposing Assumptions \ref{ass:consistency} and \ref{ass:positivity}. The difficulty arises because the observed distribution does not determine how the latent variable $U$ contributes to the conditional mean. The following proposition formally establishes this non-identifiability.

\begin{proposition}\label{prop:no_id}Under the full-data non-differential measurement mechanism and Assumptions \ref{ass:latent_uncon}-\ref{ass:positivity}, the conditional mean $\mu_{a,w}(x)$ is not identified.
\end{proposition}

To better understand the source of this non-identifiability, it is helpful to consider explicit representations of the joint distribution involving $W$ and $U$, which we now examine in our discrete setting. Define the $k_w\times k_u$ conditional probability matrix $P(W\mid U)$ as,
\begin{align*}
    P(W\mid U):=\begin{pmatrix}
        p(W=1\mid U=1) & p(W=1\mid U=2) & \dots & p(W=1\mid U=k_u)\\
        p(W=2\mid U=1) & p(W=2\mid U=2) & \dots & p(W=2\mid U=k_u)\\
        \vdots & \vdots & \ddots & \vdots\\
        p(W=k_w\mid U=1) & p(W=k_w\mid U=2) & \dots & p(W=k_w\mid U=k_u)
    \end{pmatrix}.
\end{align*}
We call this the measurement mechanism matrix, since the probability distribution describes how $W$ is generated according to $U$. It captures the uncertainty in the measurement process. In other words, even if $U$ were fixed, $W$ could still vary due to measurement errors. To illustrate that it is the source of non-identifiability, we first introduce two invertibility assumptions.

\begin{assumption}\label{ass:inv1}
    The matrix $P(W\mid U)$ is left-invertible.
\end{assumption}

\begin{assumption}\label{ass:inv2}
    For every $x\in\mathcal{X}$ and $a\in\{0,1\}$, define the $k_w\times k_u$ matrix
\[
P(U\mid W,a,x):=\big(p(U=j\mid W=i,A=a,X=x)\big)_{k_w\times k_u}\quad \text{ for }i =1\dots,k_w,\text{ and } j=1\dots,k_u.
\]
We assume that $P(U\mid W,a,x)$ is left-invertible.
\end{assumption}

Under the full-data non-differential measurement mechanism and Assumptions \ref{ass:latent_uncon}--\ref{ass:inv2}, as established in Section~\ref{sec:pc_mm}, knowledge of $P(W\mid U)$ enables identification of $\mu_{a,w}(x)$ from the observed distribution. This motivates our focus on the measurement mechanism matrix as a means of addressing the non-identifiability described in Proposition~\ref{prop:no_id}.
Since $P(W\mid U)$ is generally unknown, we pursue two approaches. First, without imposing additional restrictions on this matrix, we derive partial identification bounds for the coarser conditional mean $\mu_a(x)$ and the $X$-CATE to inform an $X$-based treatment regime. Second, we impose structural assumptions on $P(W\mid U)$ and use Bayesian analysis to quantify posterior uncertainty about the desired conditional mean function $\mu_{a,w}(x)$ and the $(X,W)$-CATE, informing an $(X,W)$-based treatment regime.

\section{Partial-Identification Bounds under Unrestricted Measurement Mechanisms}\label{sec:pi_proxy}

In this section, we propose a method for designing the $X$-based treatment policies that does not impose additional assumptions beyond the baseline assumptions already in place. In particular, we impose no further restrictions on the measurement mechanism matrix. Given the non-identifiability of $\mu_{a,w}(x)$, we shift our focus to the coarser causal estimand $\mu_a(x)$ for each $a\in\{0,1\}$ and $x\in\mathcal{X}$. The counterexample in the proof of Proposition~\ref{prop:no_id} also establishes that $\mu_a(x)$ is not identified in general. For $a=1$, the two observationally equivalent models yield different values of $\mu_1(x)$ after averaging $\mu_{1,w}(x)$ over the common conditional distribution of $W$ given $X=x$. We therefore develop a method that exploits $W$ as a proxy for the latent variable $U$ to partially identify this quantity, which can be later used to bound $\tau(X)$ and facilitate the construction of $X$-based regime.

Note that, to partially identify $\mu_a(x)$, it suffices to partially identify $\mathbb{E}[Y^{(a)}\mid A=1-a,X=x]$. We derive two partial identification results by assigning different causal roles to the proxy variable $W$. We first treat $W$ as an outcome-inducing confounding proxy.

\begin{assumption}[Bridge functions] \label{ass:bridgefun}
\begin{itemize}
    \item[(1)] There exists a non-negative bridge function $h$ such that almost surely
    \begin{align*}
        \mathbb{E}[Y\mid A,X,U]=\mathbb{E}[h(W,A,X)\mid A,X,U].
    \end{align*}
    \item[(2)] There exists a non-negative bridge function $q$ such that almost surely
    \begin{align*}
        \mathbb{E}[q(W,A,X)\mid A,X,U]=\frac{p(U\mid 1-A,X)}{p(U\mid A,X)}.
    \end{align*}
\end{itemize}
\end{assumption}

\begin{proposition}\label{prop:proxy_pi}
For any $a\in\{0,1\}$ and $x\in\mathcal{X}$ fixed, we have the following results.
\begin{enumerate}
    \item \textbf{Outcome-inducing confounding proxy bounds:} Under the observed-data non-differential measurement mechanism, Assumptions \ref{ass:latent_uncon}-\ref{ass:inv1}, and Assumption \ref{ass:bridgefun} (1), we have,
\begin{align*}
    L_a^{\text{out}}(x)\leq \mathbb{E}[Y^{(a)}\mid A=1-a,X=x]&\leq U_a^{\text{out}}(x),
\end{align*}
where
\begin{align*}
    L_a^{\text{out}}(x)&:=\max\Big\{\inf\mathcal{Y},\min_{w\in\{1,\dots,k_w\}}\frac{p(W=w\mid A=1-a,X=x)}{p(W=w\mid A=a,X=x)}\mathbb{E}[Y\mid A=a,X=x]\Big\},\\
    U_a^{\text{out}}(x)&:=\min\Big\{\sup\mathcal{Y},\max_{w\in\{1,\dots,k_w\}}\frac{p(W=w\mid A=1-a,X=x)}{p(W=w\mid A=a,X=x)}\mathbb{E}[Y\mid A=a,X=x]\Big\}.
\end{align*}
\item \textbf{Treatment-inducing confounding proxy bounds:} Under the observed-data non-differential measurement mechanism, Assumptions \ref{ass:latent_uncon}-\ref{ass:inv2}, and Assumption \ref{ass:bridgefun} (2) we have,
    \begin{align*}
    L_a^{\text{tr}}(x)\leq \mathbb{E}[Y^{(a)}\mid A=1-a,X=x]&\leq U_a^{\text{tr}}(x),
\end{align*}
where \begin{align*}
    L_a^{\text{tr}}(x)&:=\min_{w\in\{1,\dots,k_w\}}m_{a,w}(x),\qquad
    U_a^{\text{tr}}(x):=\max_{w\in\{1,\dots,k_w\}}m_{a,w}(x).
\end{align*}
\end{enumerate}

Combining the preceding results, we obtain the following bounds for $\mathbb{E}[Y^{(a)}\mid A=1-a,X=x]$ under the observed-data non-differential measurement mechanism and Assumptions \ref{ass:latent_uncon}-\ref{ass:bridgefun}. Let $L_a(x)=\max\{L_a^{\text{out}}(x),L_a^{\text{tr}}(x)\}$, and $U_a(x)=\min\{U_a^{\text{out}}(x),U_a^{\text{tr}}(x)\}$, then
\begin{align*}
    L_a(x)\leq \mathbb{E}[Y^{(a)}\mid A=1-a,X=x]\leq U_a(x), \quad \forall x\in\mathcal{X}\text{ and } a\in\{0,1\}.
\end{align*}
    
\end{proposition}

\begin{remark}
    Proximal causal inference often distinguishes between treatment-inducing confounding proxies, outcome-inducing confounding proxies, and variables that serve as common causes of treatment and outcome. The observed-data non-differential measurement mechanism indicates that the proxy $W$ satisfies both of the first two roles: it may be a potential cause of both the treatment and outcome, but any association with either is only through the latent confounder $U$. Therefore, distinct partial identification results can be obtained depending on whether $W$ is viewed as a treatment- or outcome-inducing confounding proxy.
\end{remark}

By marginalizing over the treatment assignment, these bounds translate directly into bounds on $\mu_a(x)$ for each $x\in\mathcal{X}$ and $a\in\{0,1\}$. Taking differences then yields bounds on the $X$-CATE $\tau(x)$.

\begin{theorem}
    \label{thm:proxy_outcome_CATE}
    Under the observed-data non-differential measurement mechanism and Assumptions \ref{ass:latent_uncon}-\ref{ass:bridgefun}, for each $x\in\mathcal{X}$ and $a\in\{0,1\}$, let $L_a(x)$ and $U_a(x)$ be defined as in Proposition \ref{prop:proxy_pi}. Set
\begin{align*}
    \underline{\mu}_a(x)&:=\E[Y\mid A=a,X=x]p(A=a\mid X=x)+L_a(x)p(A=1-a\mid X=x),\\
    \overline{\mu}_a(x)&:=\E[Y\mid A=a,X=x]p(A=a\mid X=x)+U_a(x)p(A=1-a\mid X=x).
\end{align*}
    Then,
\begin{align*}
    L(x)\leq \tau(x)\leq  U(x),
\end{align*}
where
\begin{align*}
    L(x)=\underline{\mu}_1(x)-\overline{\mu}_0(x),\quad U(x)=\overline{\mu}_1(x)-\underline{\mu}_0(x).
\end{align*}

\end{theorem}

\section{Posterior-Credible Bounds under Structured Measurement Mechanisms}\label{sec:pc_mm}
\subsection{Parameterized Measurement Mechanism}

In this section, we propose a Bayesian approach for designing the $(X,W)$-based treatment policies by introducing additional structure on the measurement mechanism matrix. Specifically, we posit a specific parametric model for the matrix $P(W\mid U)$ and adopt a Bayesian framework to obtain the posterior distribution of $\tau(x,w)$. This parametrization substantially reduces the degrees of freedom of $P(W\mid U)$---from $k_wk_u-k_u$ free parameters to a one-dimensional family. Given the value of this parameter, the $(X,W)$-CATE is recovered.

We begin by characterizing $\mu_{a,w}(x)$ in terms of the observed data and the measurement mechanism matrix. For each $x\in\mathcal{X}$ and $a\in\{0,1\}$, define the $k_w\times 1$ probability vectors $p_x:=\big(p(W=1\mid X=x),\dots,p(W=k_w\mid X=x)\big)^\top$, and $p_{a,x}:=\big(p(W=1\mid A=a,X=x),\dots,p(W=k_w\mid A=a,X=x)
\big)^\top$. Similarly, define the $k_w\times 1$ conditional-mean vectors $\boldsymbol{m}_{a,W}(x):=\big(m_{a,1}(x),\dots,m_{a,k_w}(x)\big)^\top$, and $\boldsymbol{\mu}_{a,W}(x):=\big(\mu_{a,1}(x),\dots,\mu_{a,k_w}(x)\big)^\top$.

\begin{theorem}\label{thm:id}
    Under the full-data non-differential measurement mechanism and Assumptions \ref{ass:latent_uncon}--\ref{ass:inv2}, if the measurement mechanism $P(W\mid U)$ is known, then for each $a\in\{0,1\}$ and $x\in\mathcal{X}$, the conditional mean $\mu_{a,w}(x)$ is identifiable for all $w\in\{1,\dots,k_w\}$ from the observed conditional expectations $\boldsymbol{m}_{a,W}(x)$ and the joint conditional distribution $P(W,A\mid X=x)$. Specifically,
    \begin{align}\label{eq_id}
        &\boldsymbol{\mu}_{a,W}(x) \nonumber\\
        &=\Big\{\Big[\text{diag}\big(p_x\big)^{-1}P(W\mid U)\text{diag}\big(P^{\dagger}(W\mid U)p_x\big)\Big]\Big\} \nonumber\\
    &~~~~~\cdot\Big\{\Big(\text{diag}\big(p_{a,x}\big)^{-1}P(W\mid U)\text{diag}\big(P^{\dagger}(W\mid U)p_{a,x}\big)\Big)^{\dagger}\boldsymbol{m}_{a,W}(x)\Big\},
    \end{align}
where $\text{diag}(v)$ is the diagonal matrix with diagonal $v$, $\dagger$ denotes the Moore–Penrose pseudoinverse, which coincides with the ordinary inverse for nonsingular square matrices. By taking difference of the above equation when $a=1$ and $a=0$, $\tau(x,w)$ for all $w$ is also recovered for each $x$ given.
\end{theorem}

\begin{remark}
Theorem \ref{thm:id} can be viewed as a first-moment counterpart to the results in \citep{kuroki2014measurement}. While their analysis focuses on recovering the full potential outcome distribution under measurement error, our result requires only conditions sufficient for identifying the conditional mean $\mu_{a,w}(x)$. In addition, our framework accommodates continuous outcomes and allows for the presence of observed covariates in the system.
\end{remark}

Note that if the matrix $P(W\mid U)$ is the identity matrix $I_k$ for $k_w=k_u=k$, we have,
\begin{align*}
    &\boldsymbol{\mu}_{a,W}(x)\\
    &=\Big\{\Big[\text{diag}\big(p_x\big)^{-1}I_k\text{diag}\big(I_k^{-1}p_x\big)\Big]\Big\}\Big\{\Big(\text{diag}\big(p_{a,x}\big)^{-1}I_k\text{diag}\big(I_k^{-1}p_{a,x}\big)\Big)^{-1}\boldsymbol{m}_{a,W}(x)\Big\}\\
    &=\boldsymbol{m}_{a,W}(x).
\end{align*}
In this case, the relationship between $W$ and $U$ is deterministic, with a one-to-one mapping between their states; that is, $W$ and $U$ are effectively the same variable up to relabeling. Consequently, observing $W$ is equivalent to observing $U$, and under the full-data non-differential measurement mechanism and Assumptions \ref{ass:latent_uncon}-\ref{ass:inv2}, the conditional mean $\mu_{a,w}(x)$ is identified as $m_{a,w}(x)$.

This represents an ideal scenario where $W$ is a perfect proxy of $U$. However, in practice, measurement error could arise due to limitations of the measuring instrument or inherent randomness in the data collection process. Motivated by the true/assigned disease severity category example, we propose a model such that the uncertainty in the measurement mechanism between $W$ and $U$ can be quantified by the model class parameter.

\begin{definition}[Truncated Discrete Normal Distribution]\label{def:td_normal}
    Let $\mu,\sigma\in\mathbb{R}$ with $\sigma>0$. The truncated discrete Normal distribution with location parameter $\mu$ and scale parameter $\sigma$, and support $\mathcal{K}$ is denoted $\mathcal{N}^{TD}_{\mathcal{K}}(\mu,\sigma^2)$. It is a probability distribution defined by,
    \begin{align*}
        p(X=x)=\frac{e^{-(x-\mu)^2/(2\sigma^2)}}{\sum_{y\in\mathcal{K}}e^{-(y-\mu)^2/(2\sigma^2)}}.
    \end{align*}
\end{definition}

For the supports $\mathcal{W}=\{1,\dots,k_w\}$ and $\mathcal{U}=\{1,\dots,k_u\}$ of $W$ and $U$ respectively, we assume the distribution of $W$ given $U=u$ follows the following distribution,
\begin{align*}
    W\mid U=u\sim \mathcal{N}^{TD}_{\mathcal{W}}(u,\sigma^2), \quad\text{ for }u\in\mathcal{U}.
\end{align*}
Under this model, the conditional distribution of $W$ given $U=u$ has its mode at $w=u$, and the scale parameter $\sigma^2$ controls the degree of measurement error. As $\sigma^2\downarrow 0$, $W=u$ with probability tending to one for every $u$. Accordingly, the measurement matrix converges to the identity matrix when $k_w=k_u$ and to the identity matrix with $k_w-k_u$ zero rows appended when $k_w>k_u$. As $\sigma^2$ increases, the conditional distribution of $W$ given $U=u$ becomes more diffuse and approaches the uniform distribution on $\mathcal{W}$. Consequently, $W$ provides less information about $U$, and the bounds on $\mu_{a,w}(x)$ may become less informative.

Thus, we have written each element in the measurement matrix as a function of the parameter $\sigma^2$, which we now denote as $P(W\mid U;\sigma^2)$, and further, the Equation \eqref{eq_id} can also be written as a function of $\sigma^2$ for each $w$ and $a$.

\subsection{Bayesian Inference and Quantile-Based Bounds}\label{sec:bayes_alg}

In practice, the parameter $\sigma^2$ governing the measurement mechanism is unknown and cannot be fully recovered from the observed data. We therefore adopt a Bayesian approach and place prior distributions on $\sigma^2$. We begin by parameterizing the likelihood function. Define the nuisance function vector $r=(r_1,\dots,r_{k_u})^\top$, where
\begin{align*}
    r_u(y,a,x) := p(U=u \mid Y=y, A=a, X=x), \quad u\in\{1,\dots,k_u\},
\end{align*}
Under the full-data non-differential measurement mechanism, the observed-data likelihood conditional on $(Y,A,X)$ can be written as
\begin{align*}\label{eq:obs_lkh}
p(w \mid y,a,x;\sigma^2,r)=\sum_{u=1}^{k_u} r_u(y,a,x)p(w\mid u;\sigma^2),
\end{align*}
and our inferential target is the marginal posterior distribution
\begin{align*}
p(\sigma^2 \mid w,y,a,x).
\end{align*}

The identification results in Equation~\eqref{eq_id} additionally depend on the observed-data nuisance functions $\{p(w\mid x), p(a\mid w,x), m_{a,w}(x)\}$. We also estimate these functions using a Bayesian approach. Let $\phi=(\phi_W,\phi_A,\phi_Y)$ denote the corresponding model parameters. Specifically, 
\begin{align*}
p(W=w\mid X=x;\phi_W)
\end{align*}
is modeled using a multinomial-logit regression,
\begin{align*}
p(A=1\mid W=w,X=x;\phi_A)
\end{align*}
is modeled using stratum-specific logistic regressions, and
\begin{align*}
m_{a,w}(x;\phi_Y)
:=
\mathbb{E}_{\phi_Y}[Y\mid A=a,W=w,X=x]
\end{align*}
is modeled using stratum-specific Bayesian linear regressions. A Gaussian working likelihood is used for the outcome model, with conditional mean $m_{a,w}(x;\phi_Y)$. Posterior uncertainty is therefore summarized using empirical posterior quantiles. 

The following algorithm summarizes the full Bayesian procedure. Joint posterior samples of $(\sigma^2,r,\phi)$ obtained from the algorithm induce posterior distributions for the nuisance functions, the conditional average treatment effects, and the resulting treatment regimes. 

\begin{algorithm}
\caption{Bayesian Posterior Sampling and CATE Evaluation}
\label{alg:HMC_marg}
\begin{algorithmic}[1]
\REQUIRE Observed data
$\mathcal{D}_n=\{(W_i,Y_i,A_i,X_i)\}_{i=1}^n$;
test points $\{(W_j^\ast,X_j^\ast)\}_{j=1}^{n_{\mathrm{test}}}$;
priors $\pi(\sigma^2)$, $\pi(r)$, and $\pi(\phi)$;
number of retained posterior draws $S$.

\STATE Specify Bayesian working models for
\[
p(w\mid x;\phi_W),\qquad
p(a\mid w,x;\phi_A),\qquad
m_{a,w}(x;\phi_Y).
\]

\STATE Define the posterior
\begin{align}\label{eq:pos_sr}
\Pi_{1,n}(\sigma^2,r\mid \mathcal{D}_n)
\propto
\prod_{i=1}^n
\left[
\sum_{u=1}^k
p(W_i\mid U_i=u,\sigma^2)
r_u(Y_i,A_i,X_i)
\right]\pi(r)\pi(\sigma^2),
\end{align}
\begin{align}\label{eq:pos_phi}
\Pi_{2,n}(\phi\mid\mathcal{D}_n)
\propto\prod_{i=1}^n\left[
p(W_i\mid X_i;\phi_W)
p(A_i\mid W_i,X_i;\phi_A)
p_Y(Y_i\mid A_i,W_i,X_i;\phi_Y)\right]\pi(\phi).
\end{align}

\STATE Use Hamiltonian Monte Carlo to obtain
\[
(\sigma^{2(s)},r^{(s)})
\sim
\Pi_{1,n}(\sigma^2,r\mid\mathcal{D}_n),\quad \phi^{(s)}
\sim
\Pi_{2,n}(\phi\mid\mathcal{D}_n),
\qquad
s=1,\ldots,S.
\]

\FOR{$s=1,\ldots,S$}
    \STATE For each test point $(X_j^\ast,W_j^\ast)$, evaluate
    \[
    P^{(s)}(W=w\mid X_j^\ast),\qquad
    P^{(s)}(A=a\mid W=w,X_j^\ast),\qquad
    m_{a,w}^{(s)}(X_j^\ast),
    \]
    for $a\in\{0,1\}$ and $w=1,\ldots,k_w$.

    \STATE Compute the naive CATE
    \[
    \hat\tau_{\mathrm{naive}}^{(s)}(X_j^\ast,W_j^\ast)
    =
    m^{(s)}_{1,W_j^\ast}(X_j^\ast)
    -
    m^{(s)}_{0,W_j^\ast}(X_j^\ast).
    \]

    \STATE Substitute the nuisance-function draws into the $X$-based partial identification formulas in Theorem \ref{thm:proxy_outcome_CATE} to obtain
    \[
    \hat{L}^{(s)}(X_j^\ast)
    \quad\text{and}\quad
    \hat{U}^{(s)}(X_j^\ast)
    \]

    \STATE Construct
    \[
    p(W\mid U;\sigma^{2(s)})
    \]
    and substitute it, together with the nuisance-function draws, into the $(X,W)$-based identification formula to obtain
    \[
    \hat\tau^{(s)}(X_j^\ast,W_j^\ast).
    \]
\ENDFOR

\STATE For each test point $(X_j^\ast,W_j^\ast)$, summarize
\[
\left\{
\hat\tau_{\mathrm{naive}}^{(s)}(X_j^\ast,W_j^\ast),
\hat{L}^{(s)}(X_j^\ast),
\hat{U}^{(s)}(X_j^\ast),
\hat\tau^{(s)}(X_j^\ast,W_j^\ast)
\right\}_{s=1}^S
\]
using empirical posterior distributions.

\end{algorithmic}
\end{algorithm}

\begin{remark}
    The observed-data nuisance models and the latent measurement model involve disjoint parameter blocks. Consequently, the posterior for $\phi$ does not depend on the prior placed on $\sigma^2$. The posterior distributions of the naive-$(X,W)$ and $X$-based quantities are therefore invariant to the prior specification for $\sigma^2$, up to Monte Carlo error. In contrast, the $(X,W)$-based quantities will vary across prior specifications for $\sigma^2$.
\end{remark}

The choice of a prior for $\sigma^2$ is, in principle, straightforward, as any distribution supported on the positive real line may be used; common choices include the Gamma, inverse-Gamma, and lognormal distributions. To provide a more interpretable and unified framework for prior specification, however, we instead formulate the prior through a derived quantity that directly reflects measurement quality. Specifically, define the average measurement accuracy as
\begin{align*}
A(\sigma^2):=\frac{1}{k_u}\sum_{u=1}^{k_u}P(W=u\mid U=u;\sigma^2).
\end{align*}
This quantity is the average probability of correctly measuring the latent category, assigning equal weight to each category of $U$.

We then define the normalized measurement-quality index as
\begin{align*}
Q(\sigma^2):=\frac{A(\sigma^2)-1/k_w}{1-1/k_w}.
\end{align*}
The baseline term $1/k_w$ is used because, under the truncated discrete normal measurement family, the conditional distribution of $W$ converges to the uniform distribution over its $k_w$ categories as $\sigma^2\to\infty$. Hence,
\begin{align*}
P(W=u\mid U=u;\sigma^2)\longrightarrow \frac{1}{k_w},
\end{align*}
for every $u$, so that $A(\sigma^2)\to 1/k_w$ and $Q(\sigma^2)\to 0$. Under perfect measurement, $A(\sigma^2)=1$ and $Q(\sigma^2)=1$.

By construction, $Q(\sigma^2) \in [0,1]$, where larger values indicate higher measurement quality. Under the truncated discrete normal family, $Q(\sigma^2)$ is continuous and strictly decreasing in $\sigma^2$. Therefore, the model can equivalently be parameterized in terms of $Q$ rather than $\sigma^2$.

\begin{definition}
We say that a prior $\pi_1$ over $\sigma^2$ is more optimistic than a prior $\pi_2$ if the induced distributions of $Q(\sigma^2)$ satisfy the following. 
\begin{align*}
    P_{\pi_1}\big(Q(\sigma^2)\geq t\big)\geq P_{\pi_2}\big(Q(\sigma^2)\geq t\big)
\end{align*}for all $t\in[0,1]$.
\end{definition}

To summarize key features of a prior $\pi$, we introduce the following quantities:
\begin{itemize}
\item \textbf{Optimism score:} $O(\pi) := \mathbb{E}_{\pi}[Q(\sigma^2)]$,
\item \textbf{Strength of belief:} $C(\pi) := \big(\operatorname{Var}_{\pi}(Q(\sigma^2))\big)^{-1}$.
\end{itemize}

Motivated by this parameterization, we place a prior directly on $Q$ rather than on $\sigma^2$. Specifically, we assume
\begin{align*}
Q \sim \mathrm{Beta}(\kappa m, \kappa(1 - m)),
\end{align*}
where $m \in (0,1)$ encodes the prior mean (i.e., the optimism score), and $\kappa > 0$ controls the concentration, reflecting the strength of belief. The induced prior on $\sigma^2$ is then obtained via a change of variables:
\begin{align*}
    \pi_{m,\kappa}(\sigma^2)=P_{\text{Beta}(\kappa m, \kappa(1-m))}\big(Q(\sigma^2)\big)\cdot\left| \frac{d}{d\sigma^2} Q(\sigma^2) \right|.
\end{align*}
Here, larger values of $m$ correspond to more optimistic beliefs about measurement quality, while larger values of $\kappa$ correspond to stronger prior concentration. In practice, we consider a grid of $(m,\kappa)$ values to represent a range of prior beliefs with varying levels of optimism and confidence.

The choice of prior for the nuisance parameter $r$ also requires careful consideration, as $r$ may be a flexible conditional distribution due to the presence of potentially continuous variables $Y$ and $X$ in its conditioning set. To accommodate this flexibility while avoiding restrictive parametric assumptions, we place a multinomial-logit Gaussian process (GP) prior on $r$. Specifically, write $Z=(Y,A,X)$, and let $D$ denote the dimension of $Z$. For each class in $\{1,\dots,k_u\}$, we represent the class probabilities through $k_u-1$ latent score functions $\{g_u(\cdot)\}_{u=1}^{k_u-1}$, with class $k_u$ used as the baseline. For each $u\in\{1,\dots,k_u-1\}$, we place an independent mean-zero GP prior
\begin{align*}
g_u(\cdot)\sim\mathcal{GP}\big(0,K_u(\cdot,\cdot)\big),
\end{align*}
and define the multinomial-logit mapping
\begin{align*}
    r_u(\cdot):=\begin{cases}
        \frac{\exp\big(c_u+g_u(\cdot)\big)}{1+\sum_{u=1}^{k_u-1}\exp\big(c_u+g_u(\cdot)\big)},\quad u=1,\dots,k_u-1;\\
        \frac{1}{1+\sum_{u=1}^{k_u-1}\exp\big(c_u+g_u(\cdot)\big)},\quad u=k_u.
    \end{cases}
\end{align*}
where $c_u$’s are class-specific intercepts, with the baseline logit fixed at zero.

We use a squared-exponential covariance kernel with dimension-specific lengthscale parameters. Specifically, for generic inputs $z,z'\in\mathbb{R}^D$, the kernel takes the form
\begin{align*}
    K_u(z,z')=\alpha_u^2\exp\Big(-\frac{1}{2}\sum_{d=1}^D\frac{(z_d-z'_d)^2}{\rho^2_{d}}\Big),
\end{align*}
where $\alpha_u>0$ controls the marginal scale of the latent GP. To improve simulation efficiency, we do not sample the lengthscale parameters $\rho_{d}$ in the posterior computation. Instead, we fix them using a prespecified empirical rule based on the scale of each component of $Z$. In particular, for each dimension $d$, we set
\begin{align*}
\rho_d
=
\begin{cases}
\dfrac{\max_i Z_{id}-\min_i Z_{id}}{2}, & \text{if } \max_i Z_{id}>\min_i Z_{id},\\
1, & \text{otherwise}.
\end{cases}
\end{align*}
This choice stabilizes computation by reducing the number of hyperparameters that need to be sampled, while still allowing the GP prior to account for the empirical scale of each component of the conditioning set $(Y,A,X)$.

Conditional on $\{g_u\}_{u=1}^{k_u-1}$ and $\{c_u\}_{u=1}^{k_u-1}$, the latent class assignments satisfy,
\begin{align*}
    U\mid (Y,A,X)\sim\text{Categorical}\big(r_1(Y,A,X),\dots,r_{k_u}(Y,A,X)\big).
\end{align*}
Therefore, under the multinomial-logit GP prior for $r$, the posterior computation only requires specifying priors for the finite-dimensional parameters $\{c_u,\alpha_u\}_{u=1}^{k_u-1}$, while the lengthscale vector $\rho=(\rho_1,\dots,\rho_D)$ is fixed by the rule above.

The algorithm yields posterior draws of the measurement parameter, the observed-data nuisance functions, and the resulting causal quantities. In particular, each draw of $\sigma^2$ and the nuisance parameters induces corresponding draws of $\mu_{a,w}(x)$ for $a\in\{0,1\}$, the naive CATE, the lower and upper endpoints of the $X$-based CATE interval, and the $(X,W)$-based CATE. Empirical posterior quantiles can then be used to summarize uncertainty in these quantities and to construct quantile-based bounds. The resulting bounds support the construction of naive, $X$-based, and $(X,W)$-based treatment regimes while accounting for uncertainty from both the measurement mechanism and the estimated nuisance functions.

\subsection{Local Non-Identification of $\sigma^2$ under Model Specifications}
Although we specify a parametric truncated discrete normal model for $P(W\mid U;\sigma^2)$ and place a flexible Gaussian process prior on $r(z)=P(U\mid Z=z)$, these model specifications do not necessarily ensure point identification of $\sigma^2$ from the observed data. In particular, the flexibility of $r(z)$ may allow different values of $\sigma^2$ to induce the same observed-data distribution. The following theorem establishes sufficient conditions under which $\sigma^2$ remains locally non-identified.

\begin{theorem}\label{thm:local_nonid_sigma2}
    Suppose that, for each $\sigma^2$ in a neighborhood of the true value
$\sigma_0^2$, the measurement mechanism matrix
\begin{align*}
    P(W\mid U;
    \sigma^2)
\end{align*}
is invertible and continuously differentiable in $\sigma^2$. Suppose also that
the true latent class probability function
\begin{align*}
    r_{0,u}(z)=p(U=u\mid Z=z)
\end{align*}
is uniformly bounded away from the boundary of the simplex, that is, there exists some $\epsilon>0$,
\begin{align*}
    \inf_{z\in\mathcal Z}\min_{u=1,\dots,k_u} r_{0,u}(z)>\epsilon.
\end{align*}
Then $\sigma^2$ is not locally point identified. In particular, there exists
$\delta>0$ such that every
\begin{align*}
    \tilde{\sigma}^2\in
    (\sigma_0^2-\delta,\sigma_0^2+\delta)
\end{align*}
is observationally equivalent to $\sigma_0^2$.
\end{theorem}

\section{Theoretical Properties of the Modular Bayesian Procedure}\label{sec:modular_bayesian}
We examine the modular construction in Algorithm \ref{alg:HMC_marg}, establish stability of its CATE evaluation, and give sufficient conditions for approximation to a coherent Bayesian CATE posterior.

\subsection{Modular Construction and Compatibility}
Algorithm \ref{alg:HMC_marg} defines a distribution induced by separately fitted posterior modules, rather than an established full joint posterior under the latent-variable model. For notation simplification in Section \ref{sec:modular_bayesian} only, write $M_{\sigma^2}=P(W\mid U;\sigma^2)$, and $b_{\sigma^2,r}(y,a,x)=M_{\sigma^2}r(y,a,x)$. Conditional on the observed data $\mathcal{D}_n$, the distribution of paired posterior draws is
\begin{align}\label{eq:dist_paired_post}
    \widetilde{\Pi}_n(\sigma^2,r,\phi\mid \mathcal{D}_n)=\Pi_{1,n}(\sigma^2,r\mid \mathcal{D}_n)\Pi_{2,n}(\phi\mid \mathcal{D}_n),
\end{align}
where the two factors are the posteriors defined in Equations \eqref{eq:pos_sr} and \eqref{eq:pos_phi} from Algorithm \ref{alg:HMC_marg}. This product construction is the modular approximation adopted in our method. We pair draws from the two separately fitted posterior modules independently conditional on $\mathcal{D}_n$. It does not assert that the full joint posterior under the latent-variable model has this factorization.

Let $f(w,y,a\mid x;\phi)$ denote the observed-data joint density specified by $\phi$. Joint compatibility between this observed-data model and the conditional model $b_{\sigma^2,r}$ specified by $(\sigma^2,r)$ requires that both imply the same conditional distribution of $W$ given $(Y,A,X)$. Thus, almost everywhere on the support where the marginal density of $(Y,A)$ given $X$ is positive, they must satisfy
\begin{align}\label{eq:compatibility}
    \frac{f(w,y,a\mid x;\phi)}{\sum_{v=1}^{k_w}f(v,y,a\mid x;\phi)}=\left[b_{\sigma^2,r}(y,a,x)\right]_w.
\end{align}
Independent pairing does not impose this equality. We therefore interpret the procedure as a modular Bayesian approximation. Modular Bayesian analysis is a broad strategy for controlling how information is shared between model components during inference. It keeps some components partly separate, for example by preventing a potentially misspecified component from influencing inference in another component, and can also simplify computation \citep{bayarri2009modularization}. Our method follows this strategy by fitting the measurement and observed-data modules separately and combining their draws when evaluating the CATE. The fitted parameters in either module do not feed back into the posterior calculation for the other module.

Bayesian composite-likelihood methods offer a related but distinct approach to simplifying inference. They replace a difficult full likelihood with a possibly weighted product of tractable marginal or conditional likelihood contributions and combine that product with a prior \citep{pauli2011bayesian,ribatet2012bayesian}. For example, a product of pairwise marginal densities may replace a high-dimensional joint density. The defining feature of this approach is the construction of a substitute likelihood. Modularization concerns how the model components are fitted and how information passes between them. These ideas can be used together, but modularization does not require a composite likelihood, and using a composite likelihood does not necessarily involve separately fitted posterior modules.

Modular Bayesian analysis provides the direct motivation for our construction. We cite composite-likelihood methods as a broader example of making Bayesian computation tractable through simpler likelihood components. The product form in \eqref{eq:dist_paired_post} does not by itself identify Algorithm \ref{alg:HMC_marg} with the specific procedures in those papers. We do not implement their adjustments for uncertainty quantification, and their corresponding theoretical guarantees do not automatically apply to our product of posterior modules.

Correct component estimation alone is insufficient to establish correct joint uncertainty \citep{ribatet2012bayesian,pompe2026asymptotics}. This distinction, however, does not preclude accurate evaluation of our CATE identification formula, as shown below.

\begin{sloppypar}
To make the computational implications concrete, one way to enforce coherence would be to retain $b_{\sigma^2,r}$ and replace the separately fitted observed-data module with the joint model $p(w,y,a\mid x;\sigma^2,r,\eta)=[b_{\sigma^2,r}(y,a,x)]_w p(y,a\mid x;\eta)$, where $p(y,a\mid x;\eta)$ is a conditional density for $(Y,A)$ given $X$, indexed by a nuisance parameter $\eta$. The nuisance quantities entering the CATE formula would then have to be derived from this same joint law. For continuous $Y$, define
\end{sloppypar}
\begin{align*}
    J_a(w,x)=\int [b_{\sigma^2,r}(y,a,x)]_w p(y,a\mid x;\eta) dy,
\end{align*}
which is the joint conditional probability of $W=w$ and $A=a$ given $X=x$. Under this model, at conditioning values with positive denominators,
\begin{align*}
    p(W=w\mid X=x)&=\sum_{a'=1}^1 J_{a'}(w,x),\\
    p(A=a\mid W=w,X=x)&=\frac{J_a(w,x)}{\sum_{a'=1}^1 J_{a'}(w,x)},\\
    m_{a,w}(x)&=\frac{\int y[b_{\sigma^2,r}(y,a,x)]_w p(y,a\mid x;\eta) dy}{J_a(w,x)}.
\end{align*}

For the outcome-mean identity, we also require that $|Y|$ have finite conditional expectation given $W=w,A=a,X=x$ under this joint model. The probabilities $p(W=w\mid A=a,X=x)$ are similarly obtained by normalizing $J_a(w,x)$ over $w$. For discrete outcomes, sums replace the integrals.

Because $r(y,a,x)$ is modeled through a multinomial-logit Gaussian process, the integrals above generally have no closed form. Numerical quadrature or Monte Carlo integration would require evaluating $r$ at additional outcome values for each treatment and covariate value of interest. These calculations would generally need to be updated across posterior draws as $(\sigma^2,r,\eta)$ change, adding function evaluations when computing CATEs over many covariate values. This cost arises in recovering the required nuisance quantities. The observed-data likelihood under the displayed factorization can itself be evaluated directly at the observed data, conditional on $(\sigma^2,r,\eta)$. Algorithm \ref{alg:HMC_marg} instead obtains these nuisance quantities directly from the separately fitted probability and outcome-mean regressions. 

Simpler joint parameterizations may reduce this computational burden but introduce additional modeling restrictions. In joint models that couple the components, feedback from a misspecified component can also degrade inference for other components \citep{bayarri2009modularization}. The relevant comparison therefore weighs the approximation error of the modular construction against the modeling and computational costs of enforcing coherence.

\subsection{Stability of CATE Evaluation}
Let $P_0$ denote the true observed-data law, and $\|\cdot\|_1$ denote the vector $\ell_1$ norm. For a fixed $(x,w)$, let $T:(\sigma^2,\boldsymbol{\nu})\mapsto\tau(x,w)$ be the $(X,W)$-CATE map using Equation \eqref{eq_id} in Theorem \ref{thm:id}, where $\boldsymbol{\nu}$ concatenates $p_x,\{p_{a,x}, \boldsymbol{m}_{a,W}(x)\}_{a=0,1}$, and $\sigma^2$ identifies a unique measurement mechanism matrix $P(W\mid U;\sigma^2)$ using the truncated discrete normal model. Let $\boldsymbol{\nu}_\phi$ and $\boldsymbol{\nu}_0$ denote its fitted and true observed-data arguments.

To connect the two fitted modules to CATE evaluation, consider a draw $(\sigma^2,r)$ from $\Pi_{1,n}$ and an independent draw $\phi$ from $\Pi_{2,n}$, conditional on $\mathcal{D}_n$, as in Algorithm \ref{alg:HMC_marg}. The resulting calculation is $T(\sigma^2,\boldsymbol{\nu}_\phi)$. Proposition \ref{prop:stability_modular} compares it with $T(\sigma^2,\boldsymbol{\nu}_0)$, which uses the same measurement mechanism parameter $\sigma^2$ and the true observed-data inputs. This comparison isolates the error from estimating those inputs in the second module while retaining the first module's distribution of $\sigma^2$. The function $r$ enters the first module's posterior and thereby affects the marginal distribution of $\sigma^2$, but does not appear directly in the CATE formula.

\begin{sloppypar}
To state sufficient conditions explicitly, we suppress $x$ in the outcome mean vectors $\boldsymbol{m}_{a,W}$, so that $\boldsymbol{\nu}=(p_x,p_{1,x},p_{0,x},\boldsymbol{m}_{1,W},\boldsymbol{m}_{0,W})$. A superscript 0 denotes the true observed-data value, whereas the subscripts 0 and 1 in $p_{a,x}$ and $\boldsymbol{m}_{a,W}$ index treatment. For a strictly positive $k_w$-vector $p$, define
\end{sloppypar}
\begin{align*}
    B_{\sigma^2}(p):=\text{diag}(p)^{-1}M_{\sigma^2}\text{diag}(M_{\sigma^2}^\dagger p).
\end{align*}
Here $\|\cdot\|_2$ denotes the Euclidean norm or its induced matrix norm, and we use $\sigma_{\min}(\cdot)$ to denote the smallest singular value.

\begin{assumption}[Positivity, stable inversion, and posterior concentration]\label{ass:stability_modular}
    At the fixed conditioning values $(x,w)$, there exists a measurable set $\Sigma$ of measurement-parameter values, finite constants $d,b,H_M,H_m>0$, with the following properties.
    \begin{enumerate}
        \item Every coordinate of $p_x^0,p_{1,x}^0,p_{0,x}^0$ is at least $d$, and $\max_{a\in\{0,1\}}\|\boldsymbol{m}_{a,W}^0\|_2\leq H_m$.
        \item For all $\sigma^2\in\Sigma$, the measurement mechanism matrix $M_{\sigma^2}$ has full column rank and is column-stochastic, meaning that its entries are non-negative and each column sums up to one. Moreover,
        \begin{align*}
            \|M_{\sigma^2}^\dagger\|_2\leq H_M,\qquad \min_{a\in\{0,1\}}\sigma_{\min}\{B_{\sigma^2}(p_{a,x}^0)\}\geq b.
        \end{align*}
        \item The first posterior module concentrates on measurement parameters in $\Sigma$,
        \begin{align*}
            \Pi_{1,n}\{(\sigma^2,r):\sigma^2\notin\Sigma\}\rightarrow 0,
        \end{align*}
        and the second consistently estimates the observed-data inputs,
        \begin{align*}
            \Pi_{2,n}\{\phi:\|\boldsymbol{\nu}_{\phi}-\boldsymbol{\nu}_0\|_1>\delta\}\rightarrow 0 \text{ for every }\delta>0,
        \end{align*}
        both in $P_0$-probability as $n\rightarrow\infty$.
    \end{enumerate}
\end{assumption}

The inverse bounds require uniform control over $\Sigma$. The consistency condition concerns the conditional means and probability vectors entering the CATE formula.

\begin{proposition}[Stability of modular CATE evaluation]\label{prop:stability_modular}
    Under Assumption \ref{ass:stability_modular}, there exist constants $L,\delta_{\ast}>0$, independent of $\sigma^2\in\Sigma$, such that the identification formula for CATE is defined and satisfies
    \begin{align*}
        |T(\sigma^2,\boldsymbol{\nu})-T(\sigma^2,\boldsymbol{\nu}_0)|\leq L\|\boldsymbol{\nu}-\boldsymbol{\nu}_0\|_1\text{ where }\sigma^2\in\Sigma, \|\boldsymbol{\nu}-\boldsymbol{\nu}_0\|_1\leq\delta_{\ast}.
    \end{align*}
    Let $\widetilde{\Pi}_n$ be the product of the separately fitted posterior modules in \eqref{eq:dist_paired_post}, and for all subsequent posterior statements, set $T(\sigma^2,\boldsymbol{\nu})=0$ whenever the identification formula is undefined, and write $\mathbbm{1}\{E\}$ for the indicator of an event $E$. Then for every $\epsilon>0$, setting $t_\epsilon=\min\{\delta_\ast,\epsilon/(2L)\}$ gives,
    \begin{align}\label{eq:stability_modular}
        &\widetilde{\Pi}_n\{|T(\sigma^2,\boldsymbol{\nu})-T(\sigma^2,\boldsymbol{\nu}_0)|>\epsilon\}\nonumber\\
        &=\int\mathbbm{1}\{|T(\sigma^2,\boldsymbol{\nu})-T(\sigma^2,\boldsymbol{\nu}_0)|>\epsilon\}\Pi_{1,n}(d\sigma^2,dr)\Pi_{2,n}(d\phi)\\
        &\leq\Pi_{1,n}\{(\sigma^2,r):\sigma^2\notin\Sigma\}+\Pi_{2,n}\{\phi:\|\boldsymbol{\nu}_{\phi}-\boldsymbol{\nu}_0\|_1>t_\epsilon\}\rightarrow0\nonumber
    \end{align}
    in $P_0$-probability.
\end{proposition}

The proof in Appendix \ref{app:proofs} first derives the stability bound from Assumption \ref{ass:stability_modular} and then
establishes \eqref{eq:stability_modular}. The bound separates the contribution of first-module draws outside the stable set from the contribution of estimation error in the second module. Compatibility in \eqref{eq:compatibility} is not needed to control this particular discrepancy. However, the comparison keeps the same distribution of $\sigma^2$ on both sides, so it does not establish that the first module learns correct or compatible measurement mechanisms, or that the product of the modules approximates a coherent joint posterior.

Here $T(\sigma^2,\boldsymbol{\nu}_0)$ is an oracle evaluation of the identification formula. It has a causal interpretation for a particular $\sigma^2$ if that measurement mechanism admits a latent model compatible with $P_0$ and satisfying the assumptions of Theorem \ref{thm:id}. Stability alone does not establish this compatibility.

\subsection{Approximation to a Coherent CATE Posterior}

Proposition \ref{prop:stability_modular} alone does not justify interpreting the modular quantile intervals as credible intervals under a coherent latent model. We next give additional conditions under which they approximate credible intervals from a specified coherent Bayesian model.

Use the joint model introduced above as a comparison model, with latent extension
\begin{align}\label{eq:comparison_model_ext}
    p^J(u,w,y,a\mid x;\sigma^2,r,\eta)=[M_{\sigma^2}]_{wu}r_u(y,a,x)p(y,a\mid x;\eta).
\end{align}
\begin{sloppypar}
Here $[M_{\sigma^2}]_{wu}$ is the $(w,u)$ entry of $M_{\sigma^2}$, and summing over $u$ recovers the observed-data density $[b_{\sigma^2,r}(y,a,x)]_wp(y,a\mid x;\eta)$. First consider Algorithm \ref{alg:HMC_marg} with prior and hyperparameter choices fixed independently of the responses $(W_i,Y_i,A_i)$, conditional on the observed covariates $X_i$. Let $\pi_1(\sigma^2,r)$ be its first-module prior, and assign the comparison model the product prior $\pi_1(\sigma^2,r)\pi_\eta(\eta)$, where $\pi_\eta$ is the prior for $\eta$. Assume these priors are proper and yield proper posteriors. Conditional on the observed covariates, the likelihood separates into first-module likelihood and $\prod_{i=1}^np(Y_i,A_i\mid X_i;\eta)$. Consequently its posterior is
\end{sloppypar}
\begin{align}\label{eq:comparison_model_posterior}
    \Pi^J_n(\sigma^2,r,\eta\mid \mathcal{D}_n)&=\Pi_{1,n}(\sigma^2,r\mid \mathcal{D}_n)\Pi_{\eta,n}(\eta\mid \mathcal{D}_n),\nonumber\\
    \Pi_{\eta,n}(\eta\mid \mathcal{D}_n)&\propto\Big\{\prod_{i=1}^np(Y_i,A_i\mid X_i;\eta)\Big\}\pi_\eta(d\eta).
\end{align}
Thus the comparison model preserves the first posterior module exactly. Let $\boldsymbol{\nu}^J_{\sigma^2,r,\eta}$ collect its probability and outcome-mean inputs, derived from this same joint law using the integrals above. These inputs depend on $(\sigma^2,r)$ as well as $\eta$, so they are not paired independently with the measurement parameter as in Algorithm \ref{alg:HMC_marg}.

At the fixed $(x,w)$, write $\tau^M=T(\sigma^2,\boldsymbol{\nu}_\phi)$ and $\tau^J=T(\sigma^2,\boldsymbol{\nu}^J_{\sigma^2,r,\eta})$. Their conditional cumulative distribution functions are
\begin{align*}
    \widetilde{F}_n(t)=\widetilde{\Pi}_n\{\tau^M\leq t\},\qquad F_n^J(t)=\Pi^J_n\{\tau^J\leq t\}.
\end{align*}

The following assumption is a supplement of Assumption \ref{ass:stability_modular}, and ensures a causal interpretation for the comparison model and controls approximation error at the scale needed for interval probabilities.

\begin{assumption}[Conditions for the coherent posterior comparison]\label{ass:coherent_comparison}
    The comparison model satisfies the following condition.
    \begin{enumerate}
        \item Under $\pi_1$, almost surely $M_{\sigma^2}$ is column-stochastic, has strictly positive entries and full column rank, and $r$ is a strictly positive probability vector at every conditioning value $(y,a,x)$. Under $\pi_\eta$, almost surely at every $x$ in the covariate support,
        \begin{align*}
            \int p(y,a\mid x;\eta)dy>0, \text{ for } a\in\{0,1\},\text{ and }\sum_{a=0}^1\int |y|p(y,a\mid x;\eta)dy<\infty.
        \end{align*}
        For discrete outcomes, sums replace the integrals.
        \item There is a deterministic sequence $\epsilon_n\downarrow0$ with $0<\epsilon_n<\delta_\ast$, where $\delta_\ast$ is the constant in Proposition \ref{prop:stability_modular}, such that
        \begin{align*}
            \Pi_{2,n}\{\|\boldsymbol{\nu}_\phi-\boldsymbol{\nu}_0\|_1>\epsilon_n\}&\rightarrow0,\\
            \Pi^J_n\{\|\boldsymbol{\nu}_{\sigma^2,r,\eta}-\boldsymbol{\nu}_0\|_1>\epsilon_n\}&\rightarrow0.
        \end{align*}
        \item With $h_n=2L\epsilon_n$, where $L$ is the constant in Proposition \ref{prop:stability_modular}, the largest posterior probability of an interval of radius $h_n$ under the comparison model satisfies
        \begin{align*}
            \omega_n:=\sup_{t\in\mathbb{R}}\Pi^J_n\{\tau^J\in[t-h_n,t+h_n]\}\rightarrow0.
        \end{align*}
    \end{enumerate}
    All convergence statements are in $P_0$-probability as $n\rightarrow\infty$.
\end{assumption}

The first condition is imposed on the two priors separately and gives the latent model in \eqref{eq:comparison_model_ext} a full-data extension satisfying the assumptions of Theorem \ref{thm:id}. The second requires both procedures to estimate the observed-data inputs accurately. The third rules out appreciable posterior mass in intervals as narrow as the remaining approximation error. For example, it holds if the comparison CATE posterior has a density bounded by a fixed constant with $P_0$-probability tending to one. Ordinary consistency alone does not imply this condition.

The next result controls the difference between the two entire CATE distributions and then transfers the credibility level of the modular quantile interval to the coherent comparison model.

\begin{proposition}[Approximation to a coherent Bayesian CATE posterior]\label{prop:approx_coherent} Under Assumptions \ref{ass:stability_modular} and \ref{ass:coherent_comparison}, define
\begin{align*}
    R_n=\Pi_{1,n}\{\sigma^2\notin\Sigma\}+\Pi_{2,n}\{\|\boldsymbol{\nu}_\phi-\boldsymbol{\nu}_0\|_1>\epsilon_n\}+\Pi^J_n\{\|\boldsymbol{\nu}^J_{\sigma^2,r,\eta}-\boldsymbol{\nu}_0\|_1>\epsilon_n\}.
\end{align*}
Then the cumulative distribution functions satisfy
\begin{align}\label{cvg:coherent_cdf}
    \Delta_n:=\sup_{t\in\mathbb{R}}\big|\widetilde{F}_n(t)-F_n^J(t)\big|\leq R_n+\omega_n\rightarrow0.
\end{align}

For $p\in(0,1)$, define the modular quantile $\widetilde{q}_{p,n}=\inf\{t\in\mathbb{R}:\widetilde{F}_n(t)\geq p\}$. For every fixed $\alpha\in(0,1)$, let $I_n=[\widetilde{q}_{\alpha/2,n},\widetilde{q}_{1-\alpha/2,n}]$. Its posterior probability under the coherent comparison model obeys,
\begin{align}\label{cvg:coherent_cred_int}
    \big|\Pi^J_n\{\tau^J\in I_n\}-(1-\alpha)\big|\leq2\Delta_n+2\omega_n\leq2R_n+4\omega_n\rightarrow0.
\end{align}
Both limits are in $P_0$-probability. No continuity assumption on $\widetilde{F}_n$ is required. 
\end{proposition}

The proof in Appendix \ref{app:proofs} uses the shared first-module draw to compare the two CATE
evaluations, and then controls posterior probabilities at the interval endpoints. Under the additional assumptions, a nominal 95\% modular quantile interval therefore has posterior probability approaching 95\% under the specified coherent model. This strengthens the oracle comparison in Proposition \ref{prop:stability_modular} without requiring Algorithm \ref{alg:HMC_marg} to fit that joint model. The result concerns ideal posterior quantiles. Empirical quantiles computed from a finite collection of draws introduce additional Monte Carlo error.

The theoretical setup above first considers prior settings chosen independently of the observed responses. Our implementation instead selects the GP lengthscales using the observed $(Y,A,X)$, including the outcome and treatment variables whose distribution is modeled by the joint comparison. Since the lengthscales determine part of the GP prior, this is an empirical Bayes procedure, meaning that the data are used to choose part of the prior. The same posterior factorization and probability bounds apply when the comparison model uses the same selected lengthscales, provided the stated assumptions hold for the resulting distributions. Both procedures then treat these selected values as fixed when computing posterior uncertainty. Thus, for our implementation, the comparison is with an empirical Bayes CATE distribution. The fixed-prior Bayesian interpretation requires choosing these prior settings independently of the modeled responses, for example using independent data.

\section{Treatment Policy Design}\label{sec:policy_design}

In this section, we consider how our bound estimates for the CATE can be used for treatment policy design. Alternative approaches construct treatment rules by comparing bounds on the conditional means of $Y^{(1)}$ and $Y^{(0)}$. Depending on how these bounds are used, such rules may emphasize either conservative or more permissive treatment recommendations. While these strategies may be appropriate in certain contexts, our focus in this work is on treatment policy design based directly on bounds for the CATE.

\subsection{Healthcare Decision-Making Strategy}

We adopt a conservative healthcare decision principle under which treatment is recommended only when the posterior lower $\alpha$-quantile of the relevant benefit measure exceeds zero. Given the observed data $\mathcal D$, define
\begin{align*}
    q_{\alpha,L}(x;\mathcal D)
    &= Q_\alpha\{L(x)\mid\mathcal D\},\\
    q_{\alpha,\tau}(x,w;\mathcal D)
    &= Q_\alpha\{\tau(x,w)\mid\mathcal D\},\\
    q_{\alpha,\mathrm{naive}}(x,w;\mathcal D)
    &= Q_\alpha\{\tau_{\mathrm{naive}}(x,w)\mid\mathcal D\},
\end{align*}
where $Q_\alpha$ denotes the posterior $\alpha$-quantile. The corresponding healthcare regimes are
\begin{align*}
    g_{\text{H-opt-}X}(x;\mathcal D)
    &= \mathbbm{1}\{q_{\alpha,L}(x;\mathcal D)>0\},\\
    g_{\text{H-opt-}(X,W)}(x,w;\mathcal D)
    &= \mathbbm{1}\{q_{\alpha,\tau}(x,w;\mathcal D)>0\},\\
    g_{\text{naive-H}}(x,w;\mathcal D)
    &= \mathbbm{1}\{q_{\alpha,\mathrm{naive}}(x,w;\mathcal D)>0\}.
\end{align*}
In implementation, we replace each posterior quantile $q_{\alpha,\cdot}$ with its empirical estimate $\hat q_{\alpha,\cdot}$ computed from posterior draws obtained from Algorithm \ref{alg:HMC_marg}, and denote the resulting regimes by $\hat g$. Smaller values of $\alpha$ yield more conservative treatment decisions.

\begin{remark}
   Proposition \ref{prop:xw_better} implies that when the CATE is point-identified and the treatment regime is determined by the sign of the CATE, the regime $g_{\text{opt}-(X,W)}$ based on richer information will outperform the regime $g_{\text{opt}-X}$ which is based only on $X$. However, when the sign of the CATE cannot be determined, namely when the bounds for the CATE contain zero, and treatment decisions rely on a default rule (for example, the healthcare decision rule given above under which treatment is withheld with insufficient evidence of benefit), neither the $g_{\text{opt-}(X,W)}$ nor the $g_{\text{opt-}X}$ necessarily dominates the other.
\end{remark}

\subsection{Combining $X$- and $(X,W)$-Healthcare optimal Regimes}
In this section, we propose a combined treatment regime that integrates the information contained in the $X$-based and $(X,W)$-based bounds. The two regimes reflect different tradeoffs: the bounds for $\tau(x)$ require fewer assumptions but are generally wider, whereas the bounds for $\tau(x,w)$ are often more informative but rely on additional assumptions on the measurement mechanism and a prior specification for $\sigma^2$. Since neither regime uniformly dominates the other, we use the $X$-based regime as the baseline recommendation and allow the $(X,W)$-based regime to modify the recommendation only when it provides evidence that treatment is harmful.

The combined regime can be expressed as,
\begin{align*}
    g_{(X,W)\text{-comb}}(x,w):=\mathbbm{1}\{q_{\alpha,L}(x;\mathcal D)>0\}\mathbbm{1}\{q_{1-\alpha,\tau}(x,w;\mathcal D)\geq0\}.
\end{align*}
The corresponding treatment decisions are summarized in Table~\ref{table:comb_reg}. When the $X$-based regime recommends withholding treatment, the combined regime also withholds treatment. When the $X$-based regime recommends treatment, we examine the $(X,W)$-based bounds. If the upper bound for $\tau(x,w)$ is negative, indicating that treatment is harmful, treatment is withheld. Otherwise, treatment is recommended. Consequently, an inconclusive $(X,W)$-based bound does not overturn a positive recommendation from the $X$-based regime, whereas evidence of harm from the $(X,W)$-based analysis always overrides it.

\begin{center}
\begin{threeparttable}
\caption{Treatment policy design based on bound estimates}
\begin{tabular}{c|c|c|c|c}
    $\tau(x,w)$ & $\tau(x)$ & $g_{\text{H-opt-}(X,W)}$ & $g_{\text{H-opt-}X}$ & Combined Decision\\ \hline
     $q_{\alpha,\tau}(x,w)>0$ & $q_{\alpha,L}(x)>0$ & $\checkmark$ & $\checkmark$ & $\checkmark$\\
     $q_{\alpha,\tau}(x,w)>0$ & $q_{1-\alpha,U}(x)<0$ & $\checkmark$ & $\times$ & $\times$\\
     $q_{\alpha,\tau}(x,w)>0$ & $q_{\alpha,L}(x)<0<q_{1-\alpha,U}(x)$ & $\checkmark$ & $-$ & $\times$\\
     $q_{1-\alpha,\tau}(x,w)<0$ & $q_{\alpha,L}(x)>0$ & $\times$ & $\checkmark$ & $\times$\\
     $q_{1-\alpha,\tau}(x,w)<0$ & $q_{1-\alpha,U}(x)<0$ & $\times$ & $\times$ & $\times$\\
     $q_{1-\alpha,\tau}(x,w)<0$ & $q_{\alpha,L}(x)<0<q_{1-\alpha,U}(x)$ & $\times$ & $-$ & $\times$\\
     $q_{\alpha,\tau}(x,w)<0<q_{1-\alpha,\tau}(x,w)$ & $q_{\alpha,L}(x)>0$ & $-$ & $\checkmark$ & $\checkmark$\\
     $q_{\alpha,\tau}(x,w)<0<q_{1-\alpha,\tau}(x,w)$ & $q_{1-\alpha,U}(x)<0$ & $-$ & $\times$ & $\times$\\
     $q_{\alpha,\tau}(x,w)<0<q_{1-\alpha,\tau}(x,w)$ & $q_{\alpha,L}(x)<0<q_{1-\alpha,U}(x)$ & $-$ & $-$ & $\times$
\end{tabular}\label{table:comb_reg}
\begin{tablenotes}
\item $q_{1-\alpha,U}(x;\mathcal D)=Q_{1-\alpha}\{U(x)\mid\mathcal D\}$;
\item $\checkmark$, treat; $\times$, withhold treatment because the treatment is harmful; $-$, withhold treatment because the sign of the treatment effect is not identified.
\end{tablenotes}
\end{threeparttable}
\end{center}

The combined rule naturally extends to multiple prior specifications for the measurement mechanism. Let $\Theta$ denote a candidate set of priors for $\sigma^2$. Rather than averaging over these priors, we treat them as a sensitivity class and use the corresponding posterior distributions only to screen for evidence of treatment harm. The combined regime is defined by
\begin{align*}
    g_{\text{comb}}:=\mathbbm{1}\{q_{\alpha,L}(x)>0\}\prod_{\pi\in\Theta}\mathbbm{1}\{q^{\pi}_{1-\alpha,\tau}(x,w)\geq0\}.
\end{align*}

Accordingly, treatment is recommended only if the $X$-based regime recommends treatment and none of the prior-specific $(X,W)$-based analyses indicates evidence of harm. Equivalently, treatment is withheld whenever the $X$-based lower bound is non-positive or at least one prior-specific posterior upper bound for $\tau(x,w)$ is negative. Thus, the collection of prior specifications acts as a posterior-harm veto on the baseline $X$-based recommendation.

The combined regimes balance the robustness of the $X$-based bounds with the greater informativeness of the $(X,W)$-based bounds, thereby enabling treatment decisions that account for both identification uncertainty and additional structural assumptions. In this sense, it is conservative with respect to additional assumptions, rather than conservative with respect to treatment per se.

\section{Simulation Study}\label{sec:sim}
In this section, we evaluate our proposed methods for treatment policy design and compare the performances of constructed regimes with the $(X,W)$-naive regime. We consider a data generating process (DGP) which satisfies the full-data non-differential measurement mechanism and Assumptions \ref{ass:latent_uncon}-\ref{ass:inv2}, and study the performances of our proposed treatment policy designs compared with the $(X,W)$-naive regime introduced in Definition \ref{def:xw-nai}. We consider a three-dimensional covariate vector $X=(X_1,X_2,X_3)$, where $X_1$ and $X_2$ are continuous covariates independently generated from a uniform distribution with support $[-1,1]$, and $X_3$ is a binary covariate generated from a Bernoulli distribution with $p(X_3=1)=0.35$. We consider a latent confounder $U$ and its noisy measurement $W$, both taking values in $\{1,\dots,5\}$. Put $v_u=u-3$. Conditional on covariate vector $X$, $U$ is generated from a five-category categorical distribution,
\begin{align*}
    U\mid X\sim\text{Categorical}\big(p_1(X),\dots,p_5(X)\big),
\end{align*}
where
\begin{align*}
    p_u(X):=p(U=u\mid X)=\frac{\exp\{0.15(2X_3-1)v_u\}}{\sum_{j=1}^5\exp\{0.15(2X_3-1)v_j\}}.
\end{align*}

Treatment $A$ is generated according to
\begin{align*}
    p(A=1\mid X_1,X_2,X_3,U=u)=\text{expit}\big(0.1 + 0.3X_3 + 0.5v_u\big).
\end{align*}
The propensity ranges from 0.28905 to 0.80218, so all latent classes retain reasonable treatment overlap. The latent means and conditional treatment effect are
\begin{align*}
    b_0(u,x)&=16+0.25x_1-0.25x_2+3.5v_u,\\
    t(u,x)&=-0.75+0.25x_1+5.5x_3+0.5v_u,\\
    b_1(u,x)&=b_0(u,x)+t(u,x)=15.25+0.5x_1-0.25x_2+5.5x_3+4v_u.
\end{align*}

Use the fixed outcome support $[L,B]=[0,32]$ and untruncated scale $s_Y=1$. For each target mean $b=b_a(u,x)$, solve the scalar equation
\begin{align*}
    b=\zeta+s_Y\frac{\phi((L-\zeta)/s_Y)-\phi((B-\zeta)/s_Y)}{\Phi((B-\zeta)/s_Y)-\Phi((L-\zeta)/s_Y)},
\end{align*}
for $\zeta=\zeta_a(u,x)$, and draw independent potential outcomes conditional on $(X,U)$,
\begin{align*}
    Y^{(a)}\mid X,U\sim \mathcal{N}(\zeta_a(U,X),s_Y^2) \text{ truncated to }[0,32], \quad Y=Y^{(A)}.
\end{align*}

\begin{remark}
The data-generating process is designed to satisfy the assumptions underlying our theoretical framework. In particular, the construction ensures treatment positivity and that the true latent class probability is uniformly bounded away from the boundary of the probability simplex, as required for our local non-identification result. Moreover, both the outcome and treatment bridge functions are non-negative over a reasonably broad range of measurement-error levels. Meanwhile, it preserves unmeasured confounding and heterogeneous treatment effects, providing a meaningful setting for evaluating the proposed treatment regimes under varying degrees of measurement error.
\end{remark}

Finally, we consider three different mechanisms through which the noisy measurement $W$ is generated.
\begin{enumerate}
    \item \textbf{The Truncated Discrete Normal Distribution} introduced in Definition \ref{def:td_normal}, with $W\mid U=u\sim \mathcal{N}^{TD}_{\mathcal{W}}(u,\sigma^2)$, for $u\in\{1,\dots,5\}$.
    \item \textbf{The Truncated Discrete Student-$t$ Distribution}. Specifically, for a fixed degrees-of-freedom parameter $\nu_0>0$, let $W\mid U=u\sim t^{TD}_{\mathcal{W}}(u,\lambda^2;\nu_0)$, for $u\in\{1,\dots,5\}$, with probability mass function,
    \begin{align*}
        p(W=w\mid U=u)=\frac{(1+\frac{(w-u)^2}{\nu_0 \lambda^2})^{-\frac{\nu_0+1}{2}}}{\sum_{j=1}^5(1+\frac{(j-u)^2}{\nu_0 \lambda^2})^{-\frac{\nu_0+1}{2}}}.
    \end{align*}
    The parameter $\lambda^2$ serves as the sole noise-level parameter, analogous to $\sigma^2$ in the truncated discrete normal model. We fix $\nu_0=3$, which induces heavier tails and allows occasional large deviations from the true state $u$, in contrast to the light-tailed behavior of the truncated discrete normal distribution.
    \item \textbf{The Diagonal-Inflated Discrete Uniform Distribution}. Let $0<\delta\leq 1$. We define $W\mid U=u \sim \mathcal{U}_{\{1,5\}}(u,\delta)$, for $u\in\{1,\dots,5\}$, with
\begin{align*}
    p(W=w\mid U=u)=
    \begin{cases}
        \delta, & \text{if } w=u,\\[4pt]
        \dfrac{1-\delta}{4}, & \text{if } w\neq u,
    \end{cases}
\end{align*}
for $w\in\{1,\dots,5\}$. The parameter $\delta$ controls the measurement accuracy: larger values of $\delta$ correspond to more accurate measurements, while smaller values yield a more diffuse, nearly uniform distribution over incorrect states.
\end{enumerate}

To examine the performance of the proposed methods under different levels of measurement error, we consider several values of the parameters controlling the noise level
\begin{enumerate}
    \item $\sigma^2\in\{0,0.25,0.5,0.65,0.75,0.85\}$.
    \item $\lambda^2\in\{0.15,0.3,0.5,0.65,0.8\}$.
    \item $\delta\in\{0.95,0.9,0.85,0.8\}$.
\end{enumerate}

While there are three candidates for the true measurement mechanism, we assess the robustness of the truncated discrete normal model by assuming $W\mid U$ follows the Truncated Discrete Normal Distribution in estimation, regardless of the true DGP. We also consider three different sample sizes $n\in\{1000,2000,4000\}$. For all cases, the testing dataset is fixed with a sample size of 1000.

For each simulated dataset, we implement Algorithm~\ref{alg:HMC_marg}. Hamiltonian Monte Carlo is run for 10,000 iterations, with the first 5,000 iterations discarded as burn-in. The remaining 5,000 posterior draws are used to approximate the posterior distributions of the naive CATE, the lower and upper endpoints of the $X$-based CATE interval, and the $(X,W)$-based CATE at every testing point. Throughout the simulations, we consider $\alpha\in\{0.05,0.1,0.2,0.3\}$, where $\alpha$ denotes the posterior quantile level used to construct the healthcare-optimal regime. Smaller values of $\alpha$ correspond to more conservative treatment decisions, requiring stronger posterior evidence that the CATE is positive. By varying $\alpha$, we evaluate the performance of the proposed regimes across different levels of treatment conservativeness.

To assess sensitivity to prior specification, we vary the prior on the measurement quality index Q through a grid of Beta distributions. Specifically, for $Q\sim\text{Beta}(\kappa m,\kappa(1-m))$, we consider parameter pairs,
\begin{align*}
    (m,\kappa)\in\{(0.5,3),(0.5,5),(0.5,10),(0.75,3),(0.75,5),(0.75,10),(0.95,3)\}.
\end{align*}
For computational stability, the Hamiltonian Monte Carlo sampler constrains the parameter space to $\sigma^2\in[0,2]$, which covers the range of measurement mechanisms considered in our numerical experiments.

For each prior specification, we obtain posterior summaries of the $(X,W)$-CATE and construct the corresponding $(X,W)$-Healthcare optimal and combined treatment regimes. Because the naive and $X$-based CATEs depend only on the observed-data nuisance functions, their posterior distributions are largely insensitive to the prior on $Q$, apart from Monte Carlo variability.

As a benchmark, we also consider a frequentist $(X,W)$-based regime. The same posterior sampling algorithm is applied, except that an uninformative uniform prior restricted to the interval $[0,2]$ is assigned to $\sigma^2$, so that the posterior primarily serves to characterize the feasible values of the sensitivity parameter that are compatible with the observed data. Unlike the Bayesian approach, which summarizes uncertainty using posterior quantiles of the $(X,W)$-based CATE, the frequentist approach constructs bounds by taking the minimum and maximum values of the posterior draws of the $(X,W)$-based CATE at each testing point. These extrema are then used as the lower and upper bounds for the frequentist $(X,W)$-based regime.

\begin{remark}
    The posterior sampler is used purely as a computational device for constructing the frequentist comparator. Unlike the Bayesian approach, which summarizes posterior uncertainty through quantiles, the frequentist comparator takes the minimum and maximum of the sampled CATE values.

    Under Algorithm~\ref{alg:HMC_marg}, the CATE depends on both the measurement parameter $\sigma^2$ and the observed-data nuisance parameters $\phi$. Specifically, for each fixed $(x,w)$, each paired posterior draw $(\sigma^{2(s)},r^{(s)},\phi^{(s)})$ induces
    \begin{align*}
        \hat{\tau}^{(s)}(x,w)
        =T(\sigma^{2(s)},\boldsymbol{\nu}_{\phi^{(s)}}),
    \end{align*}
    where $T$ is the $(X,W)$-CATE map, and $\boldsymbol{\nu}_\phi$ collects the observed-data nuisance quantities.

    Sampling $(\sigma^2,r)$ accounts for the uncertainty associated with the measurement mechanism and its latent nuisance function, while sampling $\phi$ additionally incorporates estimation uncertainty in the observed-data nuisance functions. The frequentist comparator therefore considers variation in both components through
    \begin{align*}
        \hat L_{\text{freq}}(x,w)&=\min_{s=1,\dots,S}\hat{\tau}^{(s)}(x,w),\\
        \hat U_{\text{freq}}(x,w)&=\max_{s=1,\dots,S}\hat{\tau}^{(s)}(x,w).
    \end{align*}

    If the Markov chains adequately explore their stationary distributions, then, as the number of retained draws $S\rightarrow\infty$, these extrema converge to the infimum and supremum of the CATE under the induced posterior distribution:
    \begin{align*}
        \hat{L}_{\text{freq}}(x,w)&\rightarrow\inf_{\widetilde{\Pi}}T(\sigma^2,\boldsymbol{\nu}_\phi),\\
        \hat{U}_{\text{freq}}(x,w)&\rightarrow\sup_{\widetilde{\Pi}}T(\sigma^2,\boldsymbol{\nu}_\phi),
    \end{align*}
    where $\widetilde{\Pi}$ is the product of the two posterior modules.
    
    Thus, the frequentist comparator uses posterior sampling to approximate an envelope over the posterior-supported CATE values, accounting for uncertainty in both the measurement mechanism and the observed-data nuisance functions. Unlike posterior quantiles, these extrema do not depend on a chosen credibility level.
\end{remark}

\begin{figure}[t]
  \centering
  \includegraphics[scale=0.35]{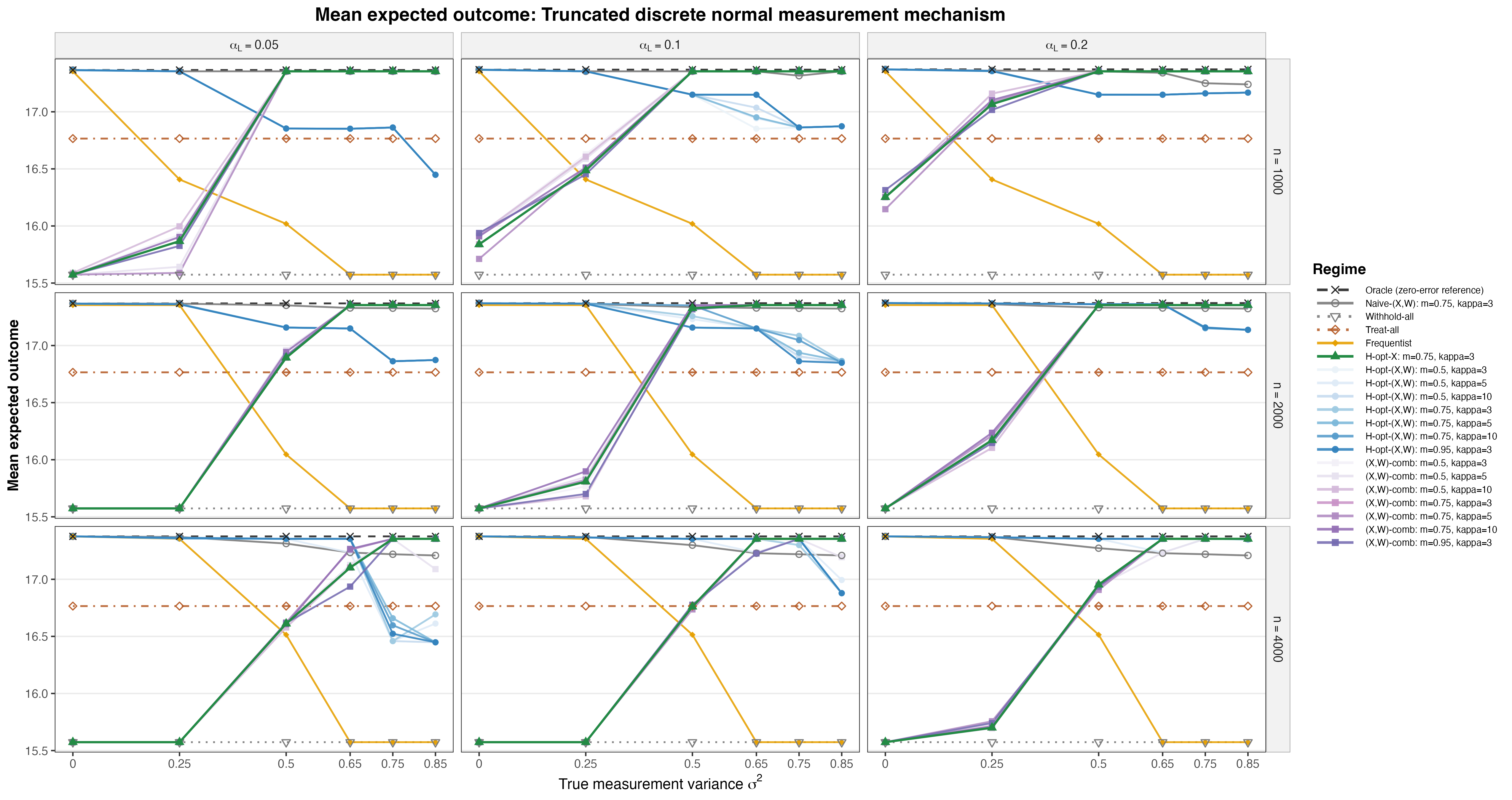}
  \caption{Mean expected outcome under the truncated discrete normal measurement mechanism for different $\sigma^2$ values and different prior specifications.}
  \label{fig:mean_exp_gaussian}
\end{figure}

\begin{figure}[t]
  \centering
  \includegraphics[scale=0.35]{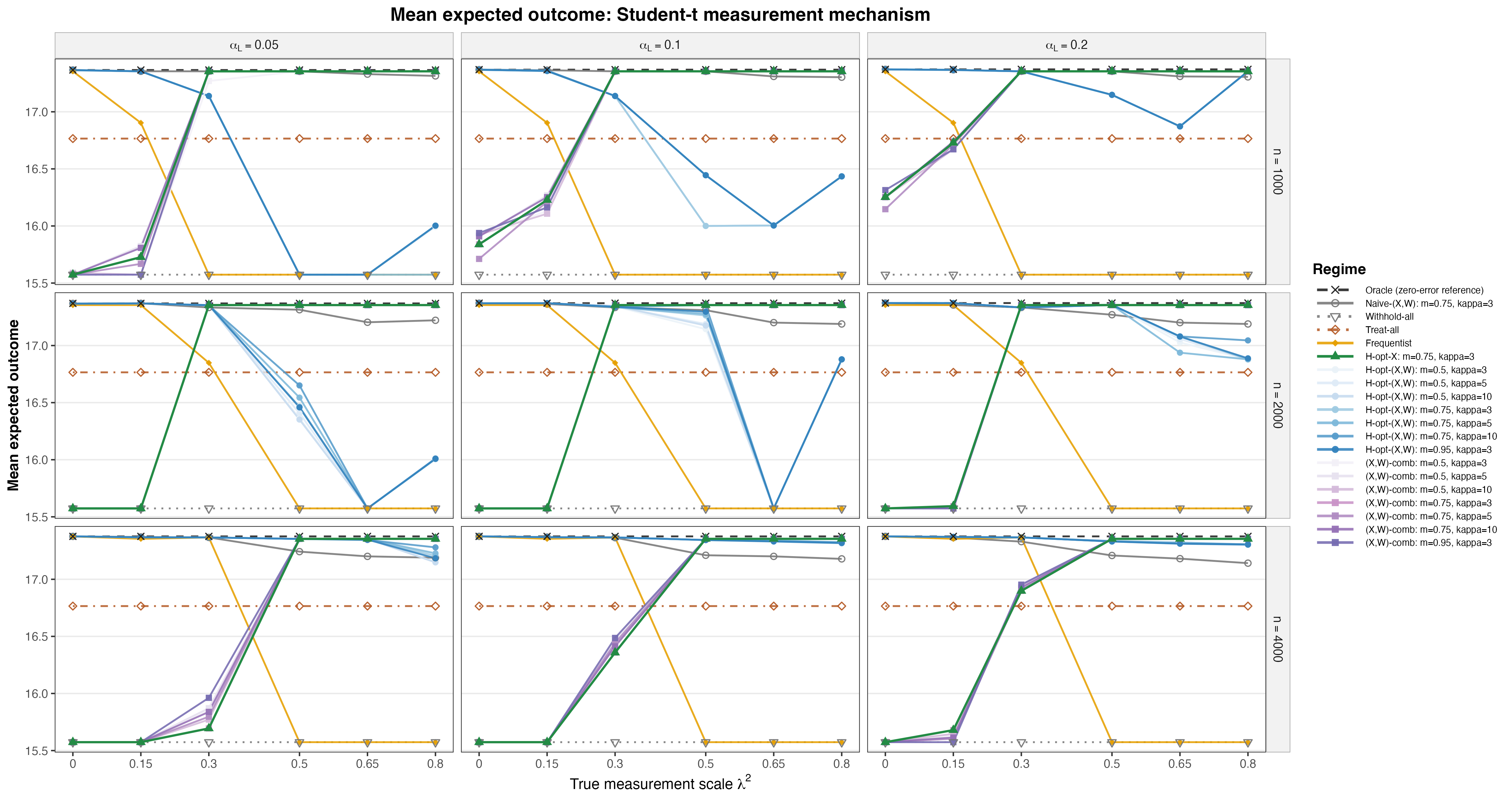}
  \caption{Mean expected outcome under the truncated discrete Student-$t$ measurement mechanism for different $\lambda^2$ values and different prior specifications.}
  \label{fig:mean_exp_stu_t}
\end{figure}

\begin{figure}[t]
  \centering
  \includegraphics[scale=0.35]{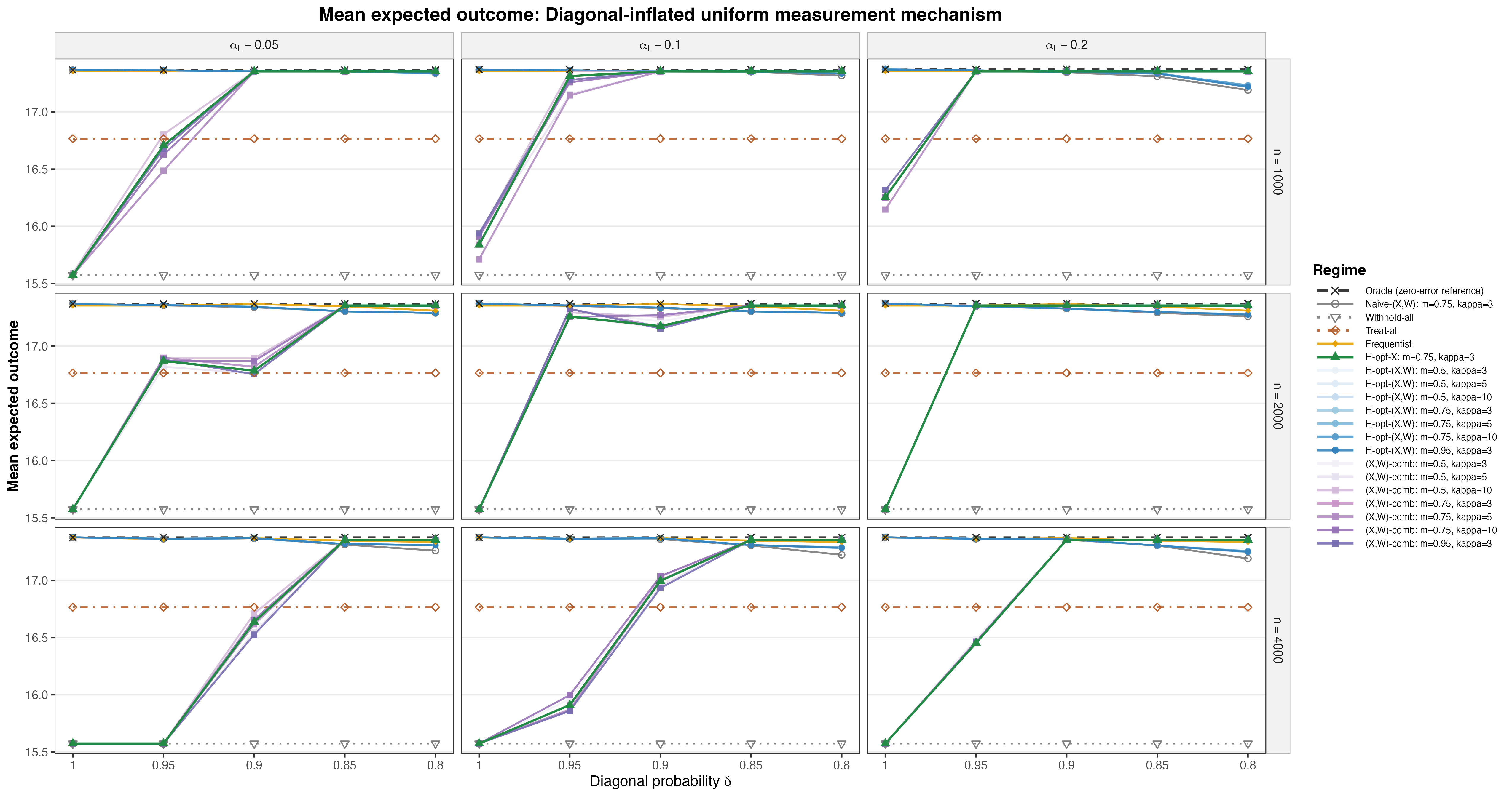}
  \caption{Mean expected outcome under the diagonal-inflated discrete uniform measurement mechanism for different $\delta$ values and different prior specifications.}
  \label{fig:mean_exp_unif}
\end{figure}

Figures~\ref{fig:mean_exp_gaussian}--\ref{fig:mean_exp_unif} reveal a systematic interaction among the conservativeness level, sample size, and measurement-error mechanism. The proposed $(X,W)$-Healthcare optimal regime achieves competitive mean expected outcomes across a broad range of measurement-error settings. When measurement error is small, $g_{\text{H-opt-}(X,W)}$ generally performs comparably to the naive regime, indicating that the proposed Bayesian procedure effectively exploits the information contained in the observed proxy. More importantly, even under substantial measurement error, $g_{\text{H-opt-}(X,W)}$ achieves high mean expected outcomes when the sample size is sufficiently large and the posterior quantile level is not overly conservative. In particular, for $n=4000$ and $\alpha\in\{0.20,0.30\}$, $g_{\text{H-opt-}(X,W)}$ frequently outperforms the naive regime and achieves outcomes comparable to the combined regimes across all three measurement-error mechanisms.

In contrast, under substantial measurement error, smaller sample sizes and more conservative posterior quantile levels can lead to noticeable deterioration in the performance of $g_{\text{H-opt-}(X,W)}$. This suggests that the effectiveness of the proposed regime depends jointly on the amount of available information and the degree of conservativeness imposed by the posterior quantile criterion. The proposed $X$-Healthcare optimal regime, $g_{\text{H-opt-}X}$, provides a complementary alternative by making treatment decisions based only on the observed covariates $X$. Its mean expected outcomes remain relatively stable across measurement-error levels and are frequently competitive with, or superior to, those of the naive regime when measurement error is substantial. This suggests that excluding the potentially unreliable proxy from treatment decisions can be beneficial when the additional information provided by $W$ is limited.

The combined regimes further exploit the complementary strengths of $g_{\text{H-opt-}X}$ and $g_{\text{H-opt-}(X,W)}$. In particular, their mean expected outcomes frequently overlap with those of $g_{\text{H-opt-}X}$, as illustrated by the overlapping green, red, and purple curves. This behavior is consistent with the construction of the combined regimes, which generally follow the treatment decisions of $g_{\text{H-opt-}X}$ unless the evidence from the $(X,W)$-based regime provides sufficiently strong support for a different decision. Consequently, the combined regimes tend to retain the stable performance of $g_{\text{H-opt-}X}$ under substantial measurement error while allowing information from $W$ to influence treatment decisions when sufficiently informative. This feature is particularly advantageous in challenging configurations where $g_{\text{H-opt-}(X,W)}$ exhibits reduced performance due to limited sample information or overly conservative posterior quantile levels.

Under the truncated discrete normal mechanism, the assumed measurement model is correctly specified, whereas the Student-$t$ and discrete uniform mechanisms assess robustness to model misspecification through relatively heavier-tailed measurement distributions. Across all three mechanisms, $g_{\text{H-opt-}(X,W)}$ achieves competitive outcomes, particularly with larger sample sizes and less conservative posterior quantile levels. Under the discrete uniform mechanism, however, the frequentist comparator generally outperforms $g_{\text{H-opt-}(X,W)}$, suggesting that the latter is more sensitive to misspecification of the measurement model in this setting. Nevertheless, $g_{\text{H-opt-}X}$ and the combined regimes exhibit relatively stable performance under substantial measurement error, even when the measurement model is misspecified.

Overall, these findings demonstrate the effectiveness and robustness of the proposed Bayesian framework across different measurement-error mechanisms. The $(X,W)$-Healthcare optimal regime effectively leverages proxy information and achieves high mean expected outcomes when sufficient sample information is available, while the $X$-Healthcare optimal and combined regimes provide stable alternatives under substantial measurement error or model misspecification. Together, the proposed regimes offer complementary advantages across varying sample sizes, conservativeness levels, and measurement-error settings.

\section{Conclusion}\label{sec:conc}
Unobserved confounding poses significant challenges to treatment policy design, as causal effects may not be point identified from observed data alone. In such settings, auxiliary information, including error-prone measurements of latent confounders, can provide valuable information for guiding treatment decisions. We developed two frameworks for personalized treatment policy design in the presence of latent confounders and their observed proxies. The first framework builds on proximal causal inference and
constructs conservative bounds on the CATE under mild invertibility conditions on the measurement mechanism. The second framework adopts a modular Bayesian approach to characterize uncertainty in the measurement mechanism and nuisance functions, using posterior credible intervals for the CATE to inform treatment decisions. Through simulation studies, we evaluated the proposed treatment policies under different measurement-error settings. The results show that the $(X,W)$-based regime achieves competitive outcomes with larger sample sizes and less conservative posterior quantile levels, while the $X$-based and combined regimes exhibit more stable performance under substantial measurement error and model misspecification.

\section*{Acknowledgments}

We are grateful to Mats J. Stensrud\footnote{Department of Mathematics,
\'Ecole Polytechnique F\'ed\'erale de Lausanne, Switzerland} for the initial discussions that inspired this
study. We also appreciate his thoughtful suggestions, which helped shape
the direction of the work and broaden the scope of the analysis.

\bibliographystyle{apalike}
\bibliography{ref.bib}

\newpage

\appendices
\section{Proofs}\label{app:proofs}

\begin{proof}[Proof of Proposition \ref{prop:xw_better}]
\begin{itemize}
    \item[(a)]

    \begin{align*}
    \mathbb{E}[Y^{(g_{\text{opt}-X})}]&=\sum_x \mathbb{E}[Y^{(g_{\text{opt-}X})}\mid X=x]p(x)\\
    &=\sum_{x,w} \mathbb{E}[Y^{(g_{\text{opt-}X})}\mid X=x,W=w]p(w\mid x)p(x)\\
    &\leq \sum_{x,w}\mathbb{E}[Y^{(g_{\text{opt-}(X,W)})}\mid X=x,W=w]p(x,w)\\
    &=\mathbb{E}[Y^{(g_{\text{opt-}(X,W)})}].
    \end{align*}

\item[(b)]
    \begin{align*}
        \mathbb{E}[Y^{(g_{\text{naive-}(X,W)})}]&=\sum_x \mathbb{E}[Y^{(g_{\text{naive-}(X,W)})}\mid X=x]p(x)\\
        &=\sum_x\Big(\sum_{w}\mathbb{E}[Y^{(g_{\text{naive-}(X,W)})}\mid X=x,W=w]p(w\mid x)\Big)p(x)\\
        &\leq\sum_x\Big(\sum_{w}\mathbb{E}[Y^{(g_{\text{opt-}(X,W)})}\mid X=x,W=w]p(w\mid x)\Big)p(x)\\
        &=\sum_x \mathbb{E}[Y^{(g_{\text{opt-}(X,W)})}\mid X=x]p(x)\\
        &=\mathbb{E}[Y^{(g_{\text{opt-}(X,W)})}].
    \end{align*}
    \end{itemize}
\end{proof}

\begin{proof}[Proof of Proposition \ref{prop:no_id}:]

To establish non-identification, it suffices to construct two observationally equivalent data-generating processes $M_1$ and $M_2$, but with different values of $\E[Y^{(a)}\mid W,X]$ as a result from unobserved distributions. We consider the special case in which $X$ is degenerate (constant), so that difference in  $\E[Y^{(a)}\mid W,X]$ reduces to difference in $\E[Y^{(a)}\mid W]$.

We also consider the setting where $A,W,U$, and $Y$ are all binary, all taking values in $\{0,1\}$. 

Take $P(U=0)=P(U=1)=0.5$. Conditional on $U$, we generate $W,A, Y^{(0)}$ and $Y^{(1)}$ independently, with the following probabilities.

    \begin{align*}
    M_1:\begin{cases}
        P(W=1\mid U=1)=0.8,\\
        P(W=1\mid U=0)=0.2,\\
        P(A=1\mid U=1)=0.8,\\
        P(A=1\mid U=0)=0.2,\\
        P(Y^{(1)}=1\mid U=1)=0.6,\\
        P(Y^{(1)}=1\mid U=0)=0.2,\\
        P(Y^{(0)}=1\mid U=1)=0.7,\\
        P(Y^{(0)}=1\mid U=0)=0.3.
    \end{cases}\quad\quad M_2:\begin{cases}
        P(W=1\mid U=1)=0.9,\\
        P(W=1\mid U=0)=0.1,\\
        P(A=1\mid U=1)=0.725,\\
        P(A=1\mid U=0)=0.275,\\
        P(Y^{(1)}=1\mid U=1)=17/29,\\
        P(Y^{(1)}=1\mid U=0)=19/55,\\
        P(Y^{(0)}=1\mid U=1)=61/110,\\
        P(Y^{(0)}=1\mid U=0)=91/290.
    \end{cases}
\end{align*}

Direct calculation gives, in both models,
\begin{align*}
    \begin{pmatrix}
        P(W=0,A=0) & P(W=0,A=1)\\
        P(W=1,A=0) & P(W=1,A=1)
    \end{pmatrix}=\begin{pmatrix}
        0.34 & 0.16\\
        0.16 & 0.34
    \end{pmatrix},
\end{align*}
and 
\begin{align*}
    \begin{pmatrix}
        P(W=0,A=0,Y=1) & P(W=0,A=1,Y=1)\\
        P(W=1,A=0,Y=1) & P(W=1,A=1,Y=1)
    \end{pmatrix}=\begin{pmatrix}
        0.11 & 0.064\\
        0.08 & 0.196
    \end{pmatrix},
\end{align*}

Because of the non-differential measurement assumption, we have,
\begin{align*}
    \E[Y^{(a)}\mid W=w]=\sum_{u=0}^1\E[Y^{(a)}\mid U=u]P(U=u\mid W=w).
\end{align*}

Under $M_1$, $\E[Y^{(1)}\mid W=1]=0.52$. Under $M_2$, $\E[Y^{(1)}\mid W=1]=\frac{4483}{7975}$. So the same observed distribution corresponds to two different values of $\E[Y^{(1)}\mid W=1]$, which proves the non-identification claim.

Moreover, under the current setup, since $X$ is degenerate, the non-identification of $\E[Y^{(a)}\mid X=x]$ reduces to showing that $\E[Y^{(a)}]$ takes different values under $M_1$ and $M_2$. Indeed, under $M_1$,
\begin{align*}
    &\E[Y^{(1)}\mid W=1]=0.52\\
    &\E[Y^{(1)}\mid W=0]=0.28,
\end{align*}
whereas under $M_2$,
\begin{align*}
    &\E[Y^{(1)}\mid W=1]=\frac{4483}{7975}\\
    &\E[Y^{(1)}\mid W=0]=\frac{2947}{7975}.
\end{align*}

Since
\begin{align*}
\mathbb{E}[Y^{(a)}]&=\sum_{w\in\{0,1\}}\mathbb{E}[Y^{(a)}\mid W=w]P(W=w),
\end{align*}
and $P(W=1)=P(W=0)=1/2$ under both models, we obtain
\begin{align*}
&\mathbb{E}_{M_1}[Y^{(1)}]=0.4\neq\frac{27}{58}=\mathbb{E}_{M_2}[Y^{(1)}].
\end{align*}

The two models therefore induce different values of $\E[Y^{(1)}]$ despite inducing the same observed-data distribution. Therefore, $\E[Y^{(a)}]$ is not identified in general. Since $X$ is degenerate in this counterexample, $\E[Y^{(a)}\mid X=x]=\E[Y^{(a)}]$, and hence $\E[Y^{(a)}\mid X=x]$ is also not identified in general.
\end{proof}

\begin{proof}[Proof of Proposition \ref{prop:proxy_pi}]
\begin{enumerate} 
    \item We have,
    \begin{align*}
        &\mathbb{E}[Y^{(a)}\mid A=1-a, X=x]\\
        &=\E\big[\E[Y^{(a)}\mid A=1-a,X=x,U]\mid A=1-a, X=x\big]\\
        &=\E\big[\E[Y^{(a)}\mid A=a,X=x,U]\mid A=1-a, X=x\big] &\text{(A\ref{ass:latent_uncon})}\\
        &=\E\big[\E[Y\mid A=a,X=x,U]\mid A=1-a, X=x\big] &\text{(A\ref{ass:consistency})}\\
        &=\E\big[\E[h(W,a,x)\mid A=a,X=x,U]\mid A=1-a, X=x\big]&\text{(A\ref{ass:bridgefun}(1))}\\
        &=\E\big[\E[h(W,a,x)\mid A=1-a,X=x,U]\mid A=1-a, X=x\big] &\text{(observed-data non-differential)}\\
        &=\E[h(W,a,x)\mid A=1-a, X=x]\\
        &=\sum_w h(w,a,x)p(w\mid 1-a,x)\\
        &=\sum_w h(w,a,x)p(w\mid a,x)\frac{p(w\mid 1-a,x)}{p(w\mid a,x)}.
    \end{align*}

    We also have,
    \begin{align*}
        &\E[Y\mid A=a,X=x]\\
        &=\E\big[\E[Y\mid A=a,X=x,U]\mid A=a,X=x\big]\\
        &=\E\big[\E[h(W,a,x)\mid A=a,X=x,U]\mid A=a, X=x\big]&\text{(A\ref{ass:bridgefun}(1))}\\
        &=\E[h(W,a,x)\mid A=a,X=x]\\
        &=\sum_w h(w,a,x)p(w\mid a,x).
    \end{align*}

    Since all weights $h(w,a,x)p(w\mid a,x)$ are non-negative, we can bound the ratio $\frac{p(w\mid 1-a,x)}{p(w\mid a,x)}$ by its minimum and maximum as,
    \begin{align*}
        &\min_w\frac{p(w\mid 1-a,x)}{p(w\mid a,x)}\sum_w h(w,a,x)p(w\mid a,x)\\
        \leq &\sum_w h(w,a,x)p(w\mid a,x)\frac{p(w\mid 1-a,x)}{p(w\mid a,x)}\\
        \leq &\max_w\frac{p(w\mid 1-a,x)}{p(w\mid a,x)}\sum_w h(w,a,x)p(w\mid a,x),
    \end{align*}
    which gives
    \begin{align*}
    &\min_{w}\frac{p(w\mid 1-a,x)}{p(w\mid A=a,x)}\mathbb{E}[Y\mid A=a,X=x]\\
    \leq &\mathbb{E}[Y^{(a)}\mid A=1-a,X=x]\\
    \leq &\max_{w}\frac{p(w\mid 1-a,x)}{p(w\mid a,x)}\mathbb{E}[Y\mid A=a,X=x].
\end{align*}

Intersecting with the outcome support, we have,
\begin{align*}
    &\quad\max\Big\{\inf\mathcal{Y},\min_{w}\frac{p(W=w\mid A=1-a,X=x)}{p(W=w\mid A=a,X=x)}\mathbb{E}[Y\mid A=a,X=x]\Big\}\\
    &\leq \mathbb{E}[Y^{(a)}\mid A=1-a,X=x]\\
    &\leq \min\Big\{\sup\mathcal{Y},\max_{w}\frac{p(W=w\mid A=1-a,X=x)}{p(W=w\mid A=a,X=x)}\mathbb{E}[Y\mid A=a,X=x]\Big\}.
\end{align*}

\item We have,
\begin{align*}
        &\mathbb{E}[Y^{(a)}\mid A=1-a, X=x]\\
        &=\E\big[\E[Y^{(a)}\mid A=1-a,X=x,U]\mid A=1-a, X=x\big]\\
        &=\E\big[\E[Y^{(a)}\mid A=a,X=x,U]\mid A=1-a, X=x\big] &\text{(A\ref{ass:latent_uncon})}\\
        &=\E\big[\E[Y\mid A=a,X=x,U]\mid A=1-a, X=x\big] &\text{(A\ref{ass:consistency})}\\
        &=\sum_{y,u}yp(y\mid a,x,u)p(u\mid 1-a,x)\\
        &=\sum_{y,u}yp(y\mid a,x,u)\frac{p(u\mid 1-a,x)}{p(u\mid a,x)}p(u\mid a,x)\\
        &=\sum_{y,u,w}yp(y\mid a,x,u)q(w,a,x)p(w\mid u,a,x)p(u\mid a,x)&\text{(A\ref{ass:bridgefun}(2))}\\
        &=\sum_{y,u,w}yp(y\mid w,a,x,u)q(w,a,x)p(w\mid u,a,x)p(u\mid a,x)&\text{(observed-data non-differential)}\\
        &=\sum_{y,u,w}yq(w,a,x)p(y,w,u\mid a,x)\\
        &=\sum_{y,w}yq(w,a,x)p(y,w\mid a,x)\\
        &=\sum_{y,w}yp(y\mid w,a,x)q(w,a,x)p(w\mid a,x)\\
        &=\sum_{w}\E[Y\mid w,a,x]q(w,a,x)p(w\mid a,x)
    \end{align*}

    Since the weights $q(w,a,x)p(w\mid a,x)$ are non-negative, and we have
    \begin{align*}
        &\sum_w q(w,a,x)p(w\mid a,x)\\
        &=\sum_{u,w}q(w,a,x)p(w\mid u,a,x)p(u\mid a,x)\\
        &=\sum_{u}\frac{p(u\mid 1-a,x)}{p(u\mid a,x)}p(u\mid a,x)&\text{(A\ref{ass:bridgefun}(2))}\\
        &=\sum_u p(u\mid 1-a,x)=1.
    \end{align*}

    Therefore, $\mathbb{E}[Y^{(a)}\mid A=1-a, X=x]$ is a convex combination of $\E[Y\mid W=w,A=a,X=x]$, and we have
    \begin{align*}
    &\min_{w}\mathbb{E}[Y\mid A=a,X=x,W=w]\\
    &\leq \mathbb{E}[Y^{(a)}\mid A=1-a,X=x]\\
    &\leq \max_{w}\mathbb{E}[Y\mid A=a,X=x,W=w].
\end{align*}
\end{enumerate}
\end{proof}

\begin{proof}[Proof of Proposition \ref{prop:stability_modular}] We first derive a uniform stability bound from Assumption \ref{ass:stability_modular} and then control the probability under the product of the two posterior modules. With $e_w$ the $w$th coordinate vector in $\mathbb{R}^{k_w}$, applying Equation \eqref{eq_id} in Theorem \ref{thm:id}, for each fixed $(x,w)$ we have
\begin{align*}
    \tau(x,w)=T(\sigma^2,\boldsymbol{\nu})=e_w^\top B_{\sigma^2}(p_x)\{B_{\sigma^2}(p_{1,x})^\dagger\boldsymbol{m}_{1,W}-B_{\sigma^2}(p_{0,x})^\dagger\boldsymbol{m}_{0,W}\}.
\end{align*}

Put $h=\|\boldsymbol{\nu}-\boldsymbol{\nu}_0\|_1$ and first restrict $h\leq\min\{d/2,1\}$. For each probability-vector argument $p$, every coordinate is then at least $d/2$, and
\begin{align*}
    \|p\|_2&\leq\|p^0\|_2+\|p-p^0\|_2\leq 1+1=2,\\
    \|\text{diag}(p)^{-1}\|_2&=\Big\|\frac{1}{p}\Big\|_\infty\leq 2/d,\\
    \|\text{diag}(p)^{-1}-\text{diag}(p^0)^{-1}\|_2&=\Big\|\text{diag}\Big(\frac{1}{p}-\frac{1}{p^0}\Big)\Big\|_2=\Big\|\frac{1}{p}-\frac{1}{p^0}\Big\|_\infty\\
    &=\Big\|\frac{1}{p}\cdot \frac{1}{p^0}\cdot(p^0-p)\Big\|_\infty\\
    &\leq\frac{2}{d}\cdot\frac{1}{d}\|p-p^0\|_2\\
    &=\frac{2}{d^2}\|p-p^0\|_2.
\end{align*}

Column stochasticity implies $\|M_{\sigma^2}\|_2\leq\sqrt{k_u}$. Also, $\|M_{\sigma^2}^\dagger(p-p^0)\|_2\leq H_M\|p-p^0\|_2$. Therefore we have, uniformly over $\sigma^2$,
\begin{align*}
    \|B_{\sigma^2}(p)\|_2&=\|\text{diag}(p)^{-1}M_{\sigma^2}\text{diag}(M_{\sigma^2}^\dagger p)\|_2\leq 2/d\cdot\sqrt{k_u}\cdot H_M\|p\|_2\leq H_B\\
    \|B_{\sigma^2}(p)-B_{\sigma^2}(p^0)\|_2&=\|\text{diag}(p)^{-1}M_{\sigma^2}\text{diag}(M_{\sigma^2}^\dagger p)-\text{diag}(p^0)^{-1}M_{\sigma^2}\text{diag}(M_{\sigma^2}^\dagger p^0)\|_2\\
    &\leq\|\text{diag}(p)^{-1}M_{\sigma^2}\text{diag}(M_{\sigma^2}^\dagger p)-\text{diag}(p^0)^{-1}M_{\sigma^2}\text{diag}(M_{\sigma^2}^\dagger p)\|_2\\
    &~~~~~+\|\text{diag}(p^0)^{-1}M_{\sigma^2}\text{diag}(M_{\sigma^2}^\dagger p)-\text{diag}(p^0)^{-1}M_{\sigma^2}\text{diag}(M_{\sigma^2}^\dagger p^0)\|_2\\
    &\leq \|\text{diag}(p)^{-1}-\text{diag}(p^0)^{-1}\|_2\cdot \sqrt{k_u}\cdot H_M\cdot2\\
    &~~~~~+\|\text{diag}(p^0)^{-1}\|_2\cdot\sqrt{k_u}\cdot\|\text{diag}(M_{\sigma^2}^\dagger p)-\text{diag}(M_{\sigma^2}^\dagger p^0)\|_2\\
    &\leq\frac{2}{d^2}\|p-p^0\|_2\cdot \sqrt{k_u}\cdot H_M\cdot2+\frac{1}{d}\sqrt{k_u}\|\text{diag}(M_{\sigma^2}^\dagger (p-p^0))\|_2\\
    &\leq\frac{4}{d^2}\cdot \sqrt{k_u}\cdot H_M\|p-p^0\|_2+\frac{1}{d}\cdot\sqrt{k_u}\cdot H_M\|p-p^0\|_2\\
    &=C_B\|p-p^0\|_2,
\end{align*}
where $H_B=4H_M\sqrt{k_u}/d$ and $C_B=H_M\sqrt{k_u}(4/d^2+1/d)$ are sufficient. Choose $\delta_{\ast}\leq\min\{d/2,1,b/(2C_B)\}$. The singular-value perturbation inequality then ensures $\sigma_{\min}\big(B_{\sigma^2}(p_{a,x})\big)\geq b/2$ for $a=0,1$. Thus, these matrices retain full column rank throughout the neighborhood.

To bound their pseudoinverses, let $F=B_{\sigma^2}(p_{a,x})$ and $G=B_{\sigma^2}(p^0_{a,x})$. Both have norm at most $H_B$ and smallest singular value at least $b/2$. Set $\Lambda=4/b^2$. The identity $F^\dagger=(F^\top F)^{-1}F^\top$ and the inverse-difference identity give,
\begin{align*}
    \|F^\dagger-G^\dagger\|_2&=\|(F^\top F)^{-1}F^\top-(G^\top G)^{-1}G^\top\|_2\\
    &\leq\|(F^\top F)^{-1}F^\top-(F^\top F)^{-1}G^\top\|_2+\|(F^\top F)^{-1}G^\top-(G^\top G)^{-1}G^\top\|_2\\
    &\leq\|(F^\top F)^{-1}\|_2\|F-G\|_2+\|(F^\top F)^{-1}-(G^\top G)^{-1}\|_2\|G\|_2\\
    &\leq\|(F^\top F)^{-1}\|_2\|F-G\|_2+\|(F^\top F)^{-1}(G^\top G-F^\top F)(G^\top G)^{-1}\|_2\|G\|_2\\
    &\leq\Lambda\|F-G\|_2+\Lambda^2H_B\|F^\top F-G^\top G\|_2\\
    &=\Lambda\|F-G\|_2+\Lambda^2H_B\|F^\top F-F^\top G+F^\top G-G^\top G\|_2\\
    &\leq\Lambda\|F-G\|_2+\Lambda^2H_B(\|F^\top F-F^\top G\|_2+\|F^\top G-G^\top G\|_2)\\
    &\leq\Lambda\|F-G\|_2+\Lambda^2H_B(\|F\|_2+\|G\|_2)\|F-G\|_2\\
    &\leq\Lambda\|F-G\|_2+2\Lambda^2H_B^2\|F-G\|_2\\
    &\leq C^\dagger C_Bh,
\end{align*}
where $C^\dagger=\Lambda+2\Lambda^2H_B^2$. In addition $\|F^\dagger\|_2\leq 2/b$, $\|G^\dagger\|_2\leq 1/b$, and $\|\boldsymbol{m}_{a,W}\|_2\leq H_m+1$.

For either treatment, expand $B_{\sigma^2}(p_{x})B_{\sigma^2}(p_{a,x})^\dagger\boldsymbol{m}_{a,W}-B_{\sigma^2}(p^0_{x})B_{\sigma^2}(p^0_{a,x})^\dagger\boldsymbol{m}_{a,W}^0$ by changing its three factors one at a time,
\begin{align*}
    &\|B_{\sigma^2}(p_{x})B_{\sigma^2}(p_{a,x})^\dagger\boldsymbol{m}_{a,W}-B_{\sigma^2}(p^0_{x})B_{\sigma^2}(p^0_{a,x})^\dagger\boldsymbol{m}_{a,W}^0\|_2\\
    &\leq\|B_{\sigma^2}(p_{x})B_{\sigma^2}(p_{a,x})^\dagger\boldsymbol{m}_{a,W}-B_{\sigma^2}(p_{x})B_{\sigma^2}(p^0_{a,x})^\dagger\boldsymbol{m}^0_{a,W}\|_2\\
    &~~~~+\|B_{\sigma^2}(p_{x})B_{\sigma^2}(p^0_{a,x})^\dagger\boldsymbol{m}^0_{a,W}-B_{\sigma^2}(p^0_{x})B_{\sigma^2}(p^0_{a,x})^\dagger\boldsymbol{m}_{a,W}^0\|_2\\
    &\leq H_B\|B_{\sigma^2}(p_{a,x})^\dagger\boldsymbol{m}_{a,W}-B_{\sigma^2}(p^0_{a,x})^\dagger\boldsymbol{m}^0_{a,W}\|_2+\|B_{\sigma^2}(p_{x})-B_{\sigma^2}(p^0_{x})\|_2\|B_{\sigma^2}(p^0_{a,x})^\dagger\boldsymbol{m}^0_{a,W}\|_2\\
    &\leq H_B\|B_{\sigma^2}(p_{a,x})^\dagger\boldsymbol{m}_{a,W}-B_{\sigma^2}(p^0_{a,x})^\dagger\boldsymbol{m}_{a,W}+B_{\sigma^2}(p^0_{a,x})^\dagger\boldsymbol{m}_{a,W}-B_{\sigma^2}(p^0_{a,x})^\dagger\boldsymbol{m}^0_{a,W}\|_2\\
    &~~~~+C_Bh\cdot 1/b\cdot H_m\\
    &\leq H_B(\|(B_{\sigma^2}(p_{a,x})^\dagger-B_{\sigma^2}(p^0_{a,x})^\dagger)\boldsymbol{m}_{a,W}\|_2+\|B_{\sigma^2}(p^0_{a,x})^\dagger(\boldsymbol{m}_{a,W}-\boldsymbol{m}^0_{a,W})\|_2)+C_Bh\cdot 1/b\cdot H_m\\
    &\leq H_B(C^\dagger C_Bh(H_m+1)+1/b\cdot h)+C_Bh\cdot 1/b\cdot H_m\\
    &=\Big\{\frac{C_B H_m}{b}+H_B C^\dagger C_B(H_m+1)+\frac{H_B}{b}\Big\}h.
\end{align*}

Summing over $a=0,1$ and using $\|e_w\|_2=1$ established the uniform stability bound in Proposition \ref{prop:stability_modular}, with twice the bracketed constant as a valid $L$.

Now fix $\epsilon>0$ and let $t_\epsilon=\min\{\delta_\ast, \epsilon/(2L)\}$. On $\{\sigma^2\in\Sigma,\|\boldsymbol{\nu}_\phi-\boldsymbol{\nu}_0\|_1\leq t_\epsilon\}$, the stability bound just established gives $\|T(\sigma^2,\boldsymbol{\nu}_\phi)-T(\sigma^2,\boldsymbol{\nu}_0)\|_1\leq\epsilon/2$. Consequently,
\begin{align*}
    \mathbbm{1}\{|T(\sigma^2,\boldsymbol{\nu}_\phi)-T(\sigma^2,\boldsymbol{\nu}_0)|>\epsilon\}\leq\mathbbm{1}\{\sigma^2\notin\Sigma\}+\mathbbm{1}\{\|\boldsymbol{\nu}_\phi-\boldsymbol{\nu}_0\|_1> t_\epsilon\}.
\end{align*}

Integrating against $\Pi_{1,n}(\sigma^2,r\mid \mathcal{D}_n)\Pi_{2,n}(\phi\mid \mathcal{D}_n)$ gives the equality and bound in \eqref{eq:stability_modular}. The two terms on its right-hand side tend to zero by the first-module concentration and second-module consistency assumptions, respectively. The bound uses only these marginal posterior probabilities and therefore also holds for any coupling with the same marginals. Independent pairing specifies the draw distribution used in Algorithm \ref{alg:HMC_marg}.
    
\end{proof}

\begin{proof}[Proof of Proposition \ref{prop:approx_coherent}] We first verify the causal interpretation of the comparison model and its posterior factorization. Equip $X$ with any fixed marginal law on the stated covariate support. Fix a parameter triple in the support specified by Assumption \ref{ass:coherent_comparison}. For each $u,a,x$, define
\begin{align*}
    H_{ua}(x)=\int r_u(y,a,x)f_\eta(y,a\mid x)dy,\qquad h_u(x)=\sum_{a=0}^1 H_{ua}(x).
\end{align*}

The support conditions imply $H_{ua}(x)>0$ and $\sum_uh_u(x)=1$. Conditional on $X=x$, generate $U$ with probabilities $h_u(x)$ and $A$ given $U=u$ with probabilities $H_{ua}(x)/h_u(x)$. Given $(U=u,X=x)$, generate the two potential outcomes independently of $A$, with respective densities
\begin{align*}
    g_{ua}(y\mid x)=\frac{r_u(y,a,x)f_\eta(y,a\mid x)}{H_{ua}(x)},\qquad a=0,1.
\end{align*}

Their conditional joint distribution may, for example, be the product of these densities. Generate $W$ using $M_{\sigma^2}(\cdot,u)$ independently of $(X,A,Y^{(0)},Y^{(1)})$ given $U=u$, and set $Y= Y^{(A)}$. This construction produces exactly \eqref{eq:comparison_model_ext}, satisfies latent unconfoundedness, consistency, treatment positivity, and the full-data non-differential measurement mechanism, and has finite conditional outcome means.

To check the rank conditions, define the column vectors
\begin{align*}
    H_a(x)=\big(H_{1a}(x),\dots,H_{k_ua}(x)\big)^\top, \quad J_a(x)=\big(J_a(1,x),\dots,J_a(k_w,x)\big)^\top.
\end{align*}
Then $J_a(x)=M_{\sigma^2}H_a(x)$. All coordinates of these vectors are positive, and Bayes rule gives
\begin{align*}
    P(U\mid W,a,x)=\text{diag}\big(J_a(x)\big)^{-1}M_{\sigma^2}\text{diag}\big(H_a(x)\big).
\end{align*}
The diagonal factors are invertible, so this matrix has full column rank, as does $M_{\sigma^2}$. Theorem \ref{thm:id} therefore identifies the CATE of this model as $\tau^J = T(\sigma^2,\boldsymbol{\nu}^J_{
\sigma^2,r,\eta})$.
Summing \eqref{eq:comparison_model_ext} over $u$ and multiplying over the observations yields
\begin{align*}
    \prod_{i=1}^n[b_{\sigma^2,r}(Y_i,A_i,X_i)]_{W_i}\prod_{i=1}^n f_{\eta}(Y_i,A_i\mid X_i).
\end{align*}

The product prior and proper normalizing constants imply the exact factorization \eqref{eq:comparison_model_posterior}. In particular, the comparison model and Algorithm~\ref{alg:HMC_marg} have the same $(\sigma^2,r)$ posterior. 

We now construct a joint distribution of the two CATE evaluations by drawing a shared $(\sigma^2,r)$ and independent $\phi$ and $\eta$, conditional on the data. Denote this coupling by
\begin{align*}
    \mathcal{Q}_n(\sigma^2,r,\phi,\eta\mid \mathcal{D}_n)=\Pi_{1,n}(\sigma^2,r\mid \mathcal{D}_n)\Pi_{2,n}(\phi\mid \mathcal{D}_n)\Pi_{\eta,n}(\eta\mid \mathcal{D}_n).
\end{align*}
Its $(\sigma^2,r,\phi)$ marginal is $\widetilde{\Pi}_n$ and its $(\sigma^2,r,\eta)$ marginal is $\Pi_n^J$. When $\sigma^2\in\Sigma$ and both nuisance-input errors are at most $\epsilon_n$, Proposition \ref{prop:stability_modular} gives
\begin{align*}
    |\tau^M-\tau^J|\leq |T(\sigma^2,\boldsymbol{\nu}_\phi)-T(\sigma^2,\boldsymbol{\nu}_0)|+|T(\sigma^2,\boldsymbol{\nu}_0)-T(\sigma^2,\boldsymbol{\nu}^J_{\sigma^2,r,\eta})|\leq 2L\epsilon_n=h_n.
\end{align*}
The union bound thus implies $\mathcal{Q}_n\{|\tau^M-\tau^J|>h_n\}\leq R_n$. For every real $t$, it follows that
\begin{align*}
    F^J_n(t-h_n)-R_n\leq \widetilde{F}_n(t)\leq F^J_n(t+h_n)+R_n.
\end{align*}
Each increment of $F^J_n$ over either adjacent interval is at most $\omega_n$. Taking the supremum over $t$
proves \eqref{cvg:coherent_cdf}.

It remains to verify the quantile statement, including possible atoms. Write $F(t-)=\lim_{v\uparrow t}F(v)$ for the left limit of a cumulative distribution function. The definition of the quantile gives
\begin{align*}
    \widetilde{F}_n(\widetilde{q}_{p,n}-)\leq p\leq\widetilde{F}_n(\widetilde{q}_{p,n}).
\end{align*}
The uniform CDF bound also holds for left limits, and every jump of $F^J_n$ is at most $\omega_n$. Consequently,
\begin{align*}
    p-\Delta_n&\leq F_n^J(\widetilde{q}_{p,n})\leq p+\Delta_n+\omega_n,\\
    p-\Delta_n-\omega_n&\leq F_n^J(\widetilde{q}_{p,n}-)\leq p+\Delta_n.
\end{align*}
Subtracting the bounds at the two endpoints yields
\begin{align*}
    1-\alpha-2\Delta_n\leq\Pi_n^J\{\tau^J\in I_n\}\leq 1-\alpha+2\Delta_n+2\omega_n,
\end{align*}
which proves \eqref{cvg:coherent_cred_int}. All these inequalities hold conditional on the data, so the data-dependent endpoints cause no additional difficulty. Finally, $R_n\rightarrow0$ follows from Assumptions \ref{ass:stability_modular} and \ref{ass:coherent_comparison}, and $\omega_n\rightarrow0$ is the last condition of Assumption \ref{ass:coherent_comparison}.
    
\end{proof}

\begin{proof}[Proof of Theorem \ref{thm:id}]
We first prove that $\E[Y^{(a)}\mid W,X]$ can be recovered from the distribution $p(w\mid u)$, and the observable conditional mean $\E[Y\mid W=w,A=a,X=x]$ and observed distributions $p(w\mid a,x)$ and $p(w\mid x)$.

For completeness, we first introduce the vector notation used in this proof.

For each $x\in\mathcal{X}$ and $a\in\{0,1\}$, define the $k_u\times 1$ probability vectors $P(U\mid x):=\big(p(U=1\mid X=x),\dots,p(U=k_u\mid X=x)\big)^\top$, and $P(U\mid a,x):=\big(p(U=1\mid A=a,X=x),\dots,p(U=k_u\mid A=a,X=x)\big)^\top$. Similarly, define the $k_w\times 1$ probability vectors $p_x:=\big(p(W=1\mid X=x),\dots,p(W=k_w\mid X=x)\big)^\top$, and $p_{a,x}:=\big(p(W=1\mid A=a,X=x),\dots,p(W=k_w\mid A=a,X=x)
\big)^\top$.

For the conditional means, define the $k_u\times 1$ vector $E(Y\mid U,a,x):=\big(E[Y\mid U=1,A=a,X=x],\dots,E[Y\mid U=k_u,A=a,X=x]\big)^\top$, and the $k_w\times 1$ vectors $\boldsymbol{m}_{a,W}(x):=\big(m_{a,1}(x),\dots,m_{a,k_w}(x)\big)^\top$, and $\boldsymbol{\mu}_{a,W}(x):=\big(\mu_{a,1}(x),\dots,\mu_{a,k_w}(x)\big)^\top$.

Under the full-data non-differential measurement mechanism, we have
\begin{align*}
p(w,y^{(a)},x\mid u)
&=p(w\mid u)p(y^{(a)},x\mid u).
\end{align*}
Therefore,
\begin{align*}
p(w,y^{(a)}\mid u,x)
&=\frac{p(w,y^{(a)},x\mid u)}{p(x\mid u)}=p(w\mid u)p(y^{(a)}\mid u,x).
\end{align*}
Moreover, we also have $p(w\mid u,x)=p(w\mid u)$. Hence,
\begin{align*}
p(w,y^{(a)}\mid u,x)
&=p(w\mid u,x)p(y^{(a)}\mid u,x),
\end{align*}
and hence
\begin{align*}
Y^{(a)}\independent W\mid U,X.
\end{align*}

Thus, by law of total probability,
\begin{align*}
    \mathbb{E}[Y^{(a)}\mid X=x,W=w]&=\sum_u \frac{p(w\mid u)p(u\mid x)}{p(w\mid x)}\mathbb{E}[Y\mid U=u,A=a,X=x],\\
    p(w\mid x)&=\sum_u p(w\mid u)p(u\mid x),\\
    p(w\mid a,x)&=\sum_u p(w\mid u)p(u\mid a,x),\\
    \mathbb{E}[Y\mid W=w,A=a,X=x]&=\sum_u \mathbb{E}[Y\mid U=u,A=a,X=x]p(u\mid w, a,x)\\
    &=\sum_u \mathbb{E}[Y\mid U=u,A=a,X=x]\frac{p(w\mid u)p(u\mid a,x)}{p(w\mid a,x)},
\end{align*}

which can be written in the matrix form as,
\begin{align}
    \boldsymbol{\mu}_{a,W}(x)&=\Big\{\text{diag}(p_x)^{-1}P(W\mid U)\text{diag}\big(P(U\mid x)\big)\Big\}\E[Y\mid U,a,x],\label{eq1}\\
    p_x&=P(W\mid U)P(U\mid x),\label{eq2}\\
    p_{a,x}&=P(W\mid U)P(U\mid a,x), \label{eq3}\\
    \boldsymbol{m}_{a,W}(x)&=P(U\mid W, a,x)E(Y\mid U,a,x)\nonumber\\
    &=\Big(\text{diag}(p_{a,x})^{-1}P(W\mid U)\text{diag}\big(P(U\mid a,x)\big)\Big)\E[Y\mid U,a,x],\label{eq4}
\end{align}
where $P(U\mid W, a,x)$ is a $k_w\times k_u$ matrix with $P(U\mid W, a,x)_{ij}=\frac{p(W=i\mid U=j)p(U=j\mid a,x)}{p(W=i\mid a,x)}$, for $i\in\{1,\dots,k_w\}$, $j\in\{1,\dots,k_u\}$, $\text{diag}(v)$ is the diagonal matrix with diagonal $v$.

For each fixed $x$ and $a \in \{0,1\}$, Equations \eqref{eq1}–\eqref{eq4} imply that the unknown components include:
\begin{itemize}
    \item the measurement mechanism $p(w \mid u)$,
    \item the distribution $p(u \mid a, x)$,
    \item the marginal distribution $p(u\mid x)$, and
    \item the outcome model $\mathbb{E}[Y \mid U=u, A=a, X=x]$.
\end{itemize}
Together, these quantities involve a total of $k_wk_u + 4k_u - 3$ free parameters, whereas Equations \eqref{eq2}–\eqref{eq4} provide $5k_w-3$ independent constraints. When $k_w=k_u=k$, the difference is $k^2-k$, which coincides with the degrees of freedom in the measurement mechanism matrix $P(W\mid U)$. 

When $k_w>k_u$, the excess degrees of freedom can be written as $(k_wk_u - k_u) - 5(k_w - k_u)$, and therefore no longer coincides with those of $P(W\mid U)$, which has $k_wk_u - k_u$ free parameters. The term $5(k_w-k_u)$ reflects the fact that the observable system is indexed by a higher-dimensional proxy $W$ rather than the latent variable $U$. These additional equations do not constitute independent identifying restrictions, but instead impose compatibility conditions, in the sense that the observed quantities must lie in the lower-dimensional subspaces induced by the latent structure. Such conditions restrict the set of admissible observed distributions but do not provide additional information for identifying the latent components.

Therefore, the central unidentified component remains the measurement mechanism $P(W\mid U)$. Our strategy is to explicitly characterize the degrees of freedom associated with $P(W\mid U)$ and impose appropriate invertibility conditions. In particular, we will require $P(W\mid U)$ to be left-invertible, which provides the algebraic basis for recovering $\boldsymbol{\mu}_{a,W}(x)$ once the measurement mechanism is specified.

Under Assumptions \ref{ass:inv1} and \ref{ass:inv2}, we can rewrite equations \eqref{eq2}, \eqref{eq3}, and \eqref{eq4} as,
\begin{align}
    P(U\mid x)&=P^{\dagger}(W\mid U)p_x, &\label{eq_ux}\\
P(U\mid a,x)&=P^{\dagger}(W\mid U)p_{a,x},&\label{eq_uax}\\
E[Y\mid U,a,x]&=P^{\dagger}(U\mid W, a,x)\boldsymbol{m}_{a,W}(x). &\label{eq_yuax}
\end{align}

Additionally, we observe that $P(U\mid W,a,x)$ can be expressed as:
\begin{align}
    P(U\mid W, a,x)=\text{diag}(p_{a,x})^{-1}P(W\mid U)\text{diag}\big(P(U\mid a,x)\big). &\label{eq_uwax}
\end{align}

These allow us to re-express the expectation of the counterfactual outcome:

\begin{align*}
    &\boldsymbol{\mu}_{a,W}(x)\\
    &\stackrel{\eqref{eq1}}{=}\Big\{\text{diag}(p_x)^{-1}P(W\mid U)\text{diag}\big(P(U\mid x)\big)\Big\}\E[Y\mid U,a,x]\\
    &\stackrel{\eqref{eq_ux}}{=}\Big\{\text{diag}(p_x)^{-1}P(W\mid U)\text{diag}\big(P^{\dagger}(W\mid U)p_x\big)\Big\}\E[Y\mid U,a,x]\\
    &\stackrel{(\ref{eq_yuax})}{=}\Big\{\text{diag}(p_x)^{-1}P(W\mid U)\text{diag}\big(P^{\dagger}(W\mid U)p_x\big)\Big\}P^{\dagger}(U\mid W, a,x)\boldsymbol{m}_{a,W}(x)\\
    &\stackrel{(\ref{eq_uwax})}{=}\Big\{\text{diag}(p_x)^{-1}P(W\mid U)\text{diag}\big(P^{\dagger}(W\mid U)p_x\big)\Big\}\\
    &~~~~~~\cdot\Big\{\Big(\text{diag}(p_{a,x})^{-1}P(W\mid U)\text{diag}\big(P(U\mid a,x)\big)\Big)^\dagger\boldsymbol{m}_{a,W}(x)\Big\}\\
    &\stackrel{(\ref{eq_uax})}{=}\Big\{\text{diag}(p_x)^{-1}P(W\mid U)\text{diag}\big(P^{\dagger}(W\mid U)p_x\big)\Big\}\\
    &~~~~~~\cdot\Big\{\Big(\text{diag}(p_{a,x})^{-1}P(W\mid U)\text{diag}\big(P^{\dagger}(W\mid U)p_{a,x}\big)\Big)^\dagger\boldsymbol{m}_{a,W}(x)\Big\}.
\end{align*}
\end{proof}

\begin{proof}[Proof of Theorem \ref{thm:local_nonid_sigma2}]
    Under the true data-generating mechanism, the observed conditional distribution
of $W$ given $Z=z$ is
\begin{align*}
    P(W\mid z)
    =
    P(W\mid U;\sigma_0^2) r_0(z),
\end{align*}
where
\begin{align*}
    r_0(z)
    =
    \big(r_{0,1}(z),\dots,r_{0,k_u}(z)\big)^\top .
\end{align*}

Now consider an alternative value
$\tilde{\sigma}^2$ close to $\sigma_0^2$. Define
\begin{align*}
    \tilde r(z)
    =
    P(W\mid U;\tilde{\sigma}^2)^{-1}
    P(W\mid U;\sigma_0^2)
    r_0(z).
\end{align*}
Because $P_{\sigma^2}(W\mid U)$ is continuously differentiable and invertible
in a neighborhood of $\sigma_0^2$,
\begin{align*}
    P(W\mid U;\tilde{\sigma}^2)^{-1}
    P(W\mid U;\sigma_0^2)
    \to I_{k_u}
    \qquad
    \text{as }
    \tilde{\sigma}^2\to\sigma_0^2.
\end{align*}
Hence $\tilde r(z)\to r_0(z)$ uniformly over $z\in\mathcal Z$.

Since
\begin{align*}
    \inf_{z\in\mathcal Z}\min_{u=1,\dots,k_u} r_{0,u}(z)>\epsilon>0.
\end{align*}
there exists $\delta>0$ such that, for all
\begin{align*}
    \tilde{\sigma}^2\in
    (\sigma_0^2-\delta,\sigma_0^2+\delta),
\end{align*}
we have
\begin{align*}
    \tilde r_u(z)\geq 0
    \qquad
    \text{for all } u=1,\dots,k_u
    \text{ and all } z\in\mathcal Z.
\end{align*}

Moreover, we also have
\begin{align*}
    \mathbf 1^\top P(W\mid U;\sigma^2)=\mathbf 1^\top .
\end{align*}
Therefore,
\begin{align*}
    \mathbf 1^\top P(W\mid U;\tilde{\sigma}^2)^{-1}=\mathbf 1^\top ,
\end{align*}
and hence
\begin{align*}
    \mathbf 1^\top \tilde r(z)
    =\mathbf 1^\top P(W\mid U;\tilde{\sigma}^2)^{-1}P(W\mid U;\sigma_0^2)r_0(z)
    =\mathbf 1^\top r_0(z)
    =1.
\end{align*}
Thus,
\begin{align*}
    \tilde r(z)\in\Delta^{k_u-1}\qquad\text{for all } z\in\mathcal Z.
\end{align*}

Finally,
\begin{align*}
    P(W\mid U;\tilde{\sigma}^2)\tilde r(z)
    &=P(W\mid U;\tilde{\sigma}^2)P(W\mid U;\tilde{\sigma}^2)^{-1}P(W\mid U;\sigma_0^2)r_0(z) \\
    &=P(W\mid U;\sigma_0^2)r_0(z) \\
    &=p(W\mid z).
\end{align*}
Therefore the alternative pair $\big(\tilde{\sigma}^2,\tilde r(z)\big)$
generates exactly the same observed conditional distribution of $W$ given
$Z=z$ as the true pair $\big(\sigma_0^2, r_0(z)\big)$. Hence every $\tilde{\sigma}^2$ sufficiently close to $\sigma_0^2$ is observationally equivalent to $\sigma_0^2$, and $\sigma^2$ is not locally point identified.
\end{proof}


\end{document}